\documentclass[a4paper, 11pt]{article}

\usepackage{a4wide}
\usepackage[normalem]{ulem}        
\usepackage{fancyhdr,fancybox}
\usepackage{amsfonts, amssymb, amsmath, ulem, amsthm, amstext, fancybox}
\usepackage{enumerate}
\usepackage{caption,graphicx,wrapfig,subfig}
\usepackage{amsthm,amsmath}

\usepackage{color}

\newcommand{\R}{\mathbb{R}}

\newcommand{\eps}{\varepsilon}

\newcommand{\D}[2]{\frac{\partial{#1}}{\partial{#2}}}

\newcommand{\recv}{\frac{1}{v}}
\newcommand{\recsigmat}{\frac{1}{\Sigma_t}}
\newcommand{\invOp}{(I +{\mathcal T})^{-1}}
\newcommand{\calX}{{\mathcal X}}

\newcommand{\calP}{{\mathcal P}}
\newcommand{\momegax}{{\mathcal M} \, \underline{\Omega}\cdot \underline{{\mathcal X}}}
\newcommand{\omegax}{\, (\underline{\Omega}\cdot \underline{{\mathcal X}}) \,}
\newcommand{\unab}{\underline{\nabla}}

\newtheoremstyle{dotless}{}{}{\itshape}{}{\bfseries}{}{ }{}

\newtheoremstyle{def}
	{0.5cm}                   
  {0.5cm}                   
  {}           							
  {}                      
  {\bfseries}  					  
  {}                      
  {\newline}        			
  {\underline{\thmname{#1} \thmnumber{#2:}} \thmnote{[#3]}}
  {}                       

\newtheorem{rem}{Remark}
\newcommand{\seqalign}[2]{\begin{subequations} #1 \begin{align} #2 \end{align}\end{subequations}}
\newcommand{\balign}[1]{\begin{align} #1 \end{align}}

\newcommand{\order}[1]{{\mathcal O} (\eps^{#1})}
\newcommand{\psix}{\Psi(\underline{x},\underline{\Omega} ,t)}
\newcommand{\phix}{\phi_0(\underline{x} ,t)}

\newcommand{\ux}{\underline{x}}
\newcommand{\uomega}{\underline{\Omega}}
\newcommand{\ul}{\underline}
\newcommand{\sigt}{\Sigma_t}
\newcommand{\siga}{\Sigma_a}
\newcommand{\sigs}{\Sigma_s}
\newcommand{\quand}{\quad \text{and} \quad}
\newcommand{\q}{\quad}
\newcommand{\tx}{\text}

\newcommand{\ds}{\displaystyle}

\newcommand{\refeq}[1]{\text{eq.~}(\ref{#1})}
\newcommand{\refeqs}[1]{\text{eqs.~}(\ref{#1})}
\newcommand{\reffig}[1]{\text{Fig.~}\ref{#1}}

\begin{document}

\title{Asymptotic Derivation and Numerical Investigation of Time-Dependent Simplified $P_N$ Equations}
\author{E. Olbrant$^{\rm a}$, E.~W.~Larsen$^{\rm b}$, M. Frank$^{\rm a}$ and B. Seibold$^{\rm c}$ \vspace{1cm} \\ $^{\rm a}$ \small{\textit{RWTH Aachen University, Department of Mathematics \& Center for Computational}} \\ \small{\textit{Engineering Science, Schinkelstrasse 2, D-52062 Aachen, Germany}} \\
$^{\rm b}$ \small{\textit{University of Michigan, Department of Nuclear Engineering and Radiological Sciences}} \\ \small{\textit{Ann Arbor, Michigan 48109-2104, USA}} \\
$^{\rm c}$ \small{\textit{Department of Mathematics, Temple University}} \\ \small{\textit{1805 North Broad Street, Philadelphia, PA 19122, USA}}
}
\date{}
\maketitle


\begin{abstract}
The steady-state simplified $P_N$ ($SP_N$) approximations to the linear Boltzmann equation have been proven to be asymptotically higher-order corrections to the diffusion equation in certain physical systems.
In this paper, we present an asymptotic analysis for the \textit{time-dependent} simplified $P_N$ equations up to $N=3$.
Additionally, $SP_N$ equations of arbitrary order are derived in an ad hoc way.
The resulting $SP_N$ equations are \textit{hyperbolic} and differ from those investigated in a previous work by some of the authors.
In two space dimensions, numerical calculations for the $P_N$ and $SP_N$ equations are performed.
We simulate neutron distributions of a moving rod and present results for a benchmark problem, known as the checkerboard problem.
The $SP_N$ equations are demonstrated to yield significantly more accurate results than diffusion approximations.
In addition, for sufficiently low values of $N$, they are shown to be more efficient than $P_N$ models of comparable cost.
\end{abstract} 
\section{Introduction}

The mathematical equation describing linear particle transport problems is the linear Boltzmann equation.
Its large dimensionality and integro-differential structure make this equation difficult to solve analytically and numerically.
Since analytic solutions of the Boltzmann equation can be constructed only for simple geometries,
a significant amount of effort has been invested to develop approximations and hence, calculate numerical solutions for realistic, multidimensional problems.
The spherical harmonics ($P_N$) equations are a standard approximation, already known from the beginning of the 20th century (first ideas for analytical solutions in \cite{Jea17} and attempts with focus on numerical computations in \cite{Cha50,Dav57}).
A drawback of the $P_N$ equations in 3-D, which has made them unattractive for practical applications, is their complicated coupling and the large number of equations, growing as $(N+1)^2$.
In the view of the low computational resources in the 1950s,
it was inevitable to come up with simpler equations for the solution of realistic problems:
Gelbard \cite{Gel60,Gel61,Gel62} therefore proposed the steady-state \textit{simplified} $P_N$ ($SP_N$) equations, which were easier to implement (because they could be written as a system of diffusion equations)
and whose system size
increases in general geometries only linearly as $(N+1)$ (versus quadratically as the $P_N$ equations do).
However, Gelbard's derivation in 3-D geometries was purely ad hoc (by taking the 1-D $P_N$ equations and replacing the 1-D spatial operators by its 3-D generalizations, i.e., gradients and divergence operators).
Due to the lack of a theoretical foundation, the $SP_N$ equations were not accepted as an approximation to the transport equation until first theoretical justifications were presented (asymptotic and variational analysis in \cite{Pom93, LarMorMcG93, TomLar96, BraLar00}).
In the framework of a Galerkin finite element method, the well-posedness of the \textit{steady-state} $SP_N$ equations is shown in \cite{ZheHan11} for $N=1,3,5,7$, where a proof of
existence and uniqueness is provided.
A detailed review of the $SP_N$ equations can be found in \cite{McC11}.

Originally intended for applications in nuclear engineering, the $SP_N$ equations are, indeed, implemented and used for neutron transport problems nowadays \cite{BecGru08, Chi11, RohGru96, PARCS}.
After first theoretical foundations for the $SP_N$ method were provided, a wide range of additional applications has been developed mainly in the past decade,
e.g., radiative cooling of glass \cite{LarThoKlaSeaGoe02, FraKlaPin11},
radiative transfer in tissue \cite{KloLar06},
fluorescence tomography \cite{KloPoe11},
design of combustion chambers for gas turbines \cite{SchSeaJanKla05, SeaFraKla04},
crystal growth of semitransparent materials \cite{BacBilRibVoi04}, and
photon and electron radiotherapy \cite{Kot07}.

The majority of previous investigations has focused on \textit{steady-state} $SP_N$ equations.
One of the ideas for deriving $SP_N$ equations is to explicitly solve for odd moments and introduce them into equations with even moments.
This is the reason why the \textit{steady-state} $SP_N$ equations reduce to a hierarchy of diffusion equations.
Keeping the time-derivatives in the odd moment equations, this procedure cannot be applied in the same way for the \textit{time-dependent} case.

To the authors' knowledge, the first formal asymptotic derivation for \textit{time-dependent} $SP_N$ equations is developed in \cite{FraKlaLar07}, and Finite Element numerical solutions of this system are computed in \cite{FraLanSch11}.
An alternative strategy for the derivation of moment methods for the time-dependent radiative transfer equation is the method of optimal prediction \cite{SeiboldFrank2009, SeiFra11}.
It turns out that this formalism yields existing moment models such as $P_N$ and diffusion closures.
Additionally, it is shown in \cite{SeiboldFrank2009} that this approach
can be used to derive variations of the parabolic $SP_N$ equations from \cite{FraKlaLar07}.

In this paper, we present an
asymptotic analysis for \textit{time-dependent} $SP_N$ equations, and we also explain how these equations can be derived in an ad hoc way.
Our analysis makes use of a
scaling which is different from \cite{FraKlaLar07} and leads to final equations which are not equivalent to those investigated in \cite{FraKlaLar07}.
We want to highlight the differences and similarities between the approach therein and the work presented here:
\begin{itemize}
\item Guided by the fact that, in \textit{steady-state}, $SP_N$ approximations are diffusion equations, the authors in \cite{FraKlaLar07} derive \textit{time-dependent} $SP_N$ equations which are \textit{parabolic} PDEs.
We follow a different approach in our asymptotic analysis and derive a system of \textit{hyperbolic} PDEs for the time-dependent $SP_N$ equations.

\item The analysis in \cite{FraKlaLar07} is performed by a parabolic scaling in which the time-derivative is scaled by $\eps^2$.
As the final $SP_N$ equations are only accurate for $\eps\approx 0$, this assertion implies that temporal changes of the solution should be very small.
In contrast, the asymptotic theory in this paper does not require a scaling of the time derivative operator.
Hence, solutions of problems with large time-derivatives should also be accurate.
However, both asymptotic theories assume that space-derivatives (scaled by $\eps$) are small.
\item The derivation in \cite{FraKlaLar07} unfolds an ambiguity of how to define the $\phi_2$ unknown.
This ambiguity is only partly captured by the introduction of a free parameter $\alpha$ in the approximate system.
Although leading to more flexibility, different choices of this parameter $\alpha$ can only give results which differ in the magnitude of $\order{6}$.
Moreover, in order to obtain a well-posed system $\alpha$ must be bounded by $0<\alpha<0.9$.
It turns out that approaching the lower or upper bound, numerical solutions of the regarding system diverge from the true solution and develop spurious shapes.
Since it is not obvious which value of $\alpha$ is ``optimal'' in some sense this issue remains unclear for practitioners. 

The analysis discussed here is similar to the asymptotic derivation of the steady-state $SP_N$ equations.
In this new analysis, there is less ambiguity and no free parameter.

\item The asymptotics in this paper and in \cite{FraKlaLar07} are
performed only up to $SP_3$.
However, as it is additionally presented here how the ad hoc derivation by Gelbard \cite{Gel60,Gel61,Gel62} can be generalized to yield exactly the same $SP_N$ equations as asymptotically derived up to $N=3$, it is straightforward to obtain higher-order $SP_N$ approximations.
\end{itemize}

The first asymptotic derivation of the time-dependent $SP_N$ equations was not developed until 2007 \cite{FraKlaLar07}. 
However, simulations with time-dependent $SP_N$ equations had been performed before.
In practice, to keep the structure of diffusion equations
for the \textit{time-dependent} generalization, some simplifications were proposed or time-derivatives were dropped in certain equations.
A different possibility, which is already included in some codes nowadays (e.g., PARCS \cite{PARCS}), was to simply add the partial time derivative to each of the steady-state equations in first-order form.
It will be shown in Section~\ref{sec:clDer} that the time-dependent $SP_3$ equations, gained heuristically in this way, can be obtained from those derived in this paper by a similarity transformation.
However, time-dependent $SP_3$ equations are also used in practice (e.g., \cite{RohGru96}) which are not equivalent to the asymptotically derived equations from Section \ref{sec:asympDer}.

This paper is organized as follows: In Section \ref{sec:clDer}, a derivation of the time-dependent 3-D $SP_3$ equations is presented which basically follows the lines of the ad hoc derivation given by Gelbard \cite{Gel60,Gel61,Gel62}.
In particular, this approach is purely ad hoc.
An asymptotic theoretical foundation is then given in Section \ref{sec:asympDer} for the time-dependent method up to order $N=3$.
It is important to stress that our asymptotic analysis in Section \ref{sec:asympDer} is only strictly valid for a homogeneous medium.
In Section \ref{sec:clDer}, heterogeneous media are considered in an ad hoc derivation.
This procedure can be generalized to $SP_N$ equations of arbitrary order, and $SP_N$ solutions are compared numerically to diffusion and $P_N$ results in Section \ref{sec:numres}.
A discussion and conclusions are given in Section \ref{sec:disc}.

\section{Classic (Ad Hoc) Derivation}
\label{sec:clDer}

In $1$-D slab geometry, the $P_N$ equations have a simple structure, and their number of unknowns is only $(N+1)$.
However, extending them to multi-dimensional geometries implies an expansion of the angular flux in spherical harmonics.
Many extra degrees of freedom are added and an additional coupling occurs.
Consequently, their original simplicity is lost and the number of equations increases quadratically in $N$.
To preserve
the pleasing form of the $1$-D slab geometry case, one can formally replace certain terms
in the $P_N$ equations in a certain way to obtain the simplified $P_N$ approximation.

The time-dependent, monoenergetic, isotropically scattering linear Boltzmann equation reads as follows:
\balign{
\frac{1}{v} \D{\Psi}{t} (\ul{x}, \ul{\Omega},t) +\ul{\Omega}\cdot \ul{\nabla} \Psi(\ul{x},\ul{\Omega},t) &+\Sigma_t(\ul{x} ,t) \Psi(\ul{x},\ul{\Omega},t) \nonumber \\
 &= \frac{\Sigma_s(\ul{x},t)}{4\pi} \int_{S^2} \Psi(\ul{x}, \ul{\Omega}',t) d\Omega' + \frac{1}{4\pi} Q(\ul{x},t), \label{intro:eq1}
 \intertext{where the parameters are}
 \sigs (\ux ,t) \, & \equiv \, \tx{scattering cross section}, \nonumber \\
 \siga (\ux ,t) \, &\equiv \, \tx{absorption cross section}, \nonumber \\
 \sigt (\ux ,t) &= \sigs (\ux ,t) +\siga (\ux ,t) \, \equiv \, \tx{total cross section}, \nonumber \\
 Q(\ux ,t) \, &\equiv \, \tx{internal source}. \nonumber
}

The angular flux $\psix$ describes the particle density at position $\ux\in\R^3$ and time $t$ traveling in direction $\uomega\in S^2$ at velocity $v$.
Penetrating the background medium, particles interact with the material, which is specified by scattering and absorption cross sections $\sigs (\ux ,t)$ and $\siga (\ux ,t)$.
However, the particles are assumed not to interact with each other.
The first term of \refeq{intro:eq1} is the temporal rate of change in $\Psi$, the second is the leakage or drift term, and the third quantifies the loss of particles due to out-scattering and absorption by the medium.
The right-hand side of \refeq{intro:eq1} characterizes the gain in particles: isotropic in-scattering processes are modeled by $\sigs (\ux ,t)$ times the integral of $\Psi$ over the unit sphere (all possible outgoing directions).
Additional particles can also be inserted into the system by an internal isotropic source $Q(\ux ,t)$.

To derive the $SP_N$ approximation to the above Boltzmann equation by a classic (ad hoc) procedure, \refeq{intro:eq1} is first restricted to planar geometry:
\begin{align}
\frac{1}{v} \D{\Psi}{t}(x,\mu,\varphi,t) &+\mu \D{\Psi}{x} (x,\mu,\varphi,t) +\Sigma_t(x,t) \Psi(x,\mu,\varphi,t) \nonumber \\
&= \frac{\Sigma_s(x,t)}{4\pi} \int_{0}^{2\pi} \int_{-1}^1 \Psi(x,\mu ',\varphi ', t) d\mu ' d\varphi ' + \frac{1}{4\pi} Q(x,t). \label{clDer:eq1}
\end{align}
Operating on \refeq{clDer:eq1} by $\int_0^{2\pi} \cdot \,\, d\varphi$ and defining the azimuthally-integrated angular flux
\[ \psi(x,\mu,t) := \int_0^{2\pi} \Psi(x,\mu,\varphi,t) d\varphi, \]
we obtain the $1$-D azimuthally-symmetric transport equation:
\begin{align}
\frac{1}{v} \D{\psi}{t} (x, \mu,t) +\mu \D{\psi}{x} (x,\mu,t) &+\Sigma_t(x,t) \psi(x,\mu,t) \nonumber \\
 &= \frac{\Sigma_s(x,t)}{2} \int_{-1}^1 \psi(x, \mu',t) d\mu' + \frac{1}{2} Q(x,t). \label{clDer:eq2}
\end{align}
Multiplying \refeq{clDer:eq2} by Legendre polynomials $P_n(\mu)$ and integrating this equation over $-1\leq\mu\leq 1$, we obtain, after using the recursion relation for $P_n(\mu)$,
\begin{align}
\frac{1}{v} \D{\phi_n}{t} (x,t) &+ \D{}{x} \left ( \frac{n+1}{2n+1} \phi_{n+1}(x,t) +\frac{n}{2n+1} \phi_{n-1} (x,t) \right ) + \Sigma_t(x,t) \phi_n(x,t) \nonumber \\
& \hspace{6cm}= \delta_{n,0} \left ( \Sigma_s(x,t) \phi_0(x,t) +Q(x,t) \right ),
\label{clDer:eq3}
\end{align}
where we defined the Legendre moments of $\psi(x,\mu,t)$ as:
\begin{align} 
\phi_n(x,t) = \int_{-1}^1 P_n(\mu) \psi(x,\mu,t) d\mu, \quad n\geq 0.
\end{align}
Eq.~(\ref{clDer:eq3}) is exact for all integers $n\geq 0$. Unfortunately, it never yields a closed system of equations; there is always one more unknown function than there are equations. The standard $P_N$ approximation is simply to set the highest Legendre moment of $\psi(x,\mu,t)$ equal to zero. In the case of the first four equations (corresponding to $n=0,1,2,3$), one sets $\phi_4(x,t) = 0$ to get the following classic planar geometry time-dependent $P_3$ equations:
\begin{subequations} \label{clDer:eq8}
\begin{align}
\frac{1}{v} \D{\phi_0}{t} (x,t) + \D{\phi_1}{x} (x,t) + \Sigma_a(x,t) \phi_0 (x,t) &= Q(x,t), \label{clDer:eq8a} \\
\frac{1}{v} \D{\phi_1}{t} (x,t) + \D{}{x} \left ( \frac{2}{3} \phi_2(x,t) +\frac{1}{3} \phi_{0}(x,t) \right ) + \Sigma_t(x,t) \phi_1(x,t) &= 0, \label{clDer:eq8b}\\
\frac{1}{v} \D{\phi_2}{t} (x,t) + \D{}{x} \left ( \frac{3}{5} \phi_3(x,t) +\frac{2}{5} \phi_1(x,t) \right ) + \Sigma_t(x,t) \phi_2(x,t) & = 0, \label{clDer:eq8c} \\
\frac{1}{v} \D{\phi_3}{t} (x,t) + \D{}{x} \left ( \frac{3}{7} \phi_2(x,t) \right ) + \Sigma_t(x,t) \phi_3(x,t) &= 0. \label{clDer:eq8d}
\end{align}
\end{subequations}
From now on, to keep the discussion simple, we will work with the specific system of \refeqs{clDer:eq8}. 
However, everything which is done in the following can be generalized to more (or fewer) than four angular moments of \refeq{clDer:eq2}.

Solving for the odd moments of \refeqs{clDer:eq8}, we divide \refeq{clDer:eq8b} and \refeq{clDer:eq8d} by $\Sigma_t(x)$ and obtain
\begin{subequations}
\begin{align}
\phi_1(x,t) &= -\frac{1}{3} (I +{\mathcal T})^{-1} {\mathcal X} (2\phi_2(x,t) +\phi_0(x,t)), \\
\phi_3(x,t) &= -\frac{1}{7} (I +{\mathcal T})^{-1} {\mathcal X} (3\phi_2(x,t)),
\end{align}
\end{subequations}
where
\begin{subequations} \label{clDer:eq9}
\begin{align}
{\mathcal T} := \frac{1}{v\Sigma_t(x,t)} \D{}{t} \quand {\mathcal X} := \frac{1}{\Sigma_t(x,t)} \D{}{x}
\end{align}
\end{subequations}
are two dimensionless operators. Introducing these expressions into the first and third of \refeqs{clDer:eq8}, we get
\seqalign{\label{clDer:eq12x}}{
\recv \D{\phi_0}{x}(x,t) &- \frac{1}{3} \D{}{x} (I +{\mathcal T})^{-1} {\mathcal X} (2\phi_2(x,t) +\phi_0(x,t)) +\Sigma_a(x,t) \phi_0(x,t) = Q(x,t), \\
\recv \D{\phi_2}{x}(x,t) &-\D{}{x} \left [ \frac{9}{35} \invOp\calX \phi_2(x,t) +\frac{2}{15} \invOp\calX (2\phi_2(x,t) +\phi_0(x,t)) \right ] \nonumber \\
& \hspace{8cm} +\sigt(x,t) \phi_2(x,t) = 0.
}
To derive the SP$_3$ equations, we formally replace the 1-D operator
\seqalign{\label{clDer:eq12}}{
\D{}{x} (I +{\mathcal T})^{-1} {\mathcal X} = \D{}{x} & \invOp \frac{1}{\sigt(x,t)} \D{}{x}
\intertext{by the 3-D operator:}
\uline{\nabla} \cdot \invOp & \frac{1}{\sigt(\ux,t)} \uline{\nabla} \\
= \D{}{x} \invOp \frac{1}{\sigt(\ux,t)} \D{}{x} +\D{}{y} \invOp & \frac{1}{\sigt(\ux,t)} \D{}{y} +\D{}{z} \invOp \frac{1}{\sigt(\ux,t)} \D{}{z}.
}
This step is purely ad hoc; yet it is exactly what is (or rather, was) done in the original (1960) derivation of the steady-state $SP_N$ equations \cite{Gel60,Gel61,Gel62}.
Eqs.~(\ref{clDer:eq12x}) then become:
\seqalign{\label{clDer:eq13}}{
\recv \D{\phi_0}{t} (\ux,t) +\siga(\ux,t) \phi_0(\ux,t) &= \unab\cdot \invOp \frac{1}{3\sigt(\ux,t)} \uline{\nabla} (\phi_0(\ux,t) +2\phi_2(\ux,t) ) +Q(\ux,t), \\
\recv \D{\phi_2}{t} (\ux,t) +\sigt(\ux,t) \phi_2(\ux,t)  &= \unab\cdot \invOp \frac{1}{3\sigt(\ux,t)} \uline{\nabla} \left(\frac{2}{5} \phi_0(\ux,t) +\frac{11}{7} \phi_2(\ux,t) \right).
}
\begin{rem}
Eqs.~(\ref{clDer:eq13}) directly reduce to the standard steady-state SP$_3$ equations when \[{\mathcal T} = \frac{1}{v\sigt(\ux,t)} \D{}{t} = 0.\]
\end{rem}
\noindent Next, we rewrite \refeqs{clDer:eq13} as a system of hyperbolic PDEs by defining
\seqalign{\label{clDer:eq14}}{
\ul{J}_0(\uline{x},t) &:= -\invOp \frac{1}{3\sigt(\uline{x},t)} \unab \bigg (\phi_0(\ux ,t) +2\phi_2(\ux, t) \bigg ), \\
\ul{J}_2(\uline{x},t) &:= -\invOp \frac{1}{3\sigt(\uline{x},t)} \unab \left( \frac{2}{5} \phi_0(\ux ,t) +\frac{11}{7} \phi_2(\ux, t) \right).
}
Then, \refeqs{clDer:eq13} can be written as:
\seqalign{\label{clDer:eq15}}{
\recv \D{\phi_0}{t} (\ux ,t) &+\unab\cdot \ul{J}_0 (\ux ,t) +\siga(\ux ,t)\phi_0 (\ux ,t) = Q(\ux,t), \\
\recv \D{\phi_2}{t} (\ux ,t) &+\unab\cdot \ul{J}_2 (\ux ,t) +\sigt(\ux ,t)\phi_2 (\ux ,t) = 0, \\
\recv \D{\ul{J}_0}{t}(\ux ,t) &+\frac{1}{3} \unab \bigg (\phi_0 (\ux ,t) +2\phi_2 (\ux ,t)  \bigg ) +\sigt(\ux ,t)\ul{J}_0 (\ux ,t)  = 0, \\
\recv \D{\ul{J}_2}{t}(\ux ,t) &+\frac{1}{3} \unab \left ( \frac{2}{5} \phi_0 (\ux ,t) +\frac{11}{7} \phi_2 (\ux ,t)  \right) +\sigt(\ux ,t)\ul{J}_2 (\ux ,t)  = 0.
}
These are time-dependent 3-D SP$_3$ equations. This coupled system of first-order PDEs is easy to discretize, spatially and temporally. 
We note that in \refeqs{clDer:eq15} all quantities (cross sections, source term and fluxes) can be functions of $\ux$ and $t$.

\begin{rem}
It is interesting to observe that the steady-state $SP_N$ equations are problematic in systems containing void regions (in which $\sigt (\ux ,t) = 0$), but \refeqs{clDer:eq15} do not have an issue with voids.
\end{rem}

\begin{rem}
Adding the partial time derivative to each of the steady-state $SP_3$ equations in first-order form formally implies \cite{McC11}:
\seqalign{\label{clDer:eq18}}{
\recv \D{\phi_0}{t} (\ux ,t) &+\unab\cdot \ul{\phi}_1 (\ux ,t) +\siga(\ux ,t) \phi_0 (\ux ,t)= Q(\ux,t), \\
\recv \D{\phi_2}{t} (\ux ,t) &+ \frac15 \unab\cdot \left (  2\ul{\phi}_1 (\ux ,t) + 3\ul{\phi}_3 (\ux ,t)  \right )+\sigt(\ux ,t) \phi_2 (\ux ,t)= 0 \\
\recv \D{\ul{\phi}_1}{t}(\ux ,t) &+\frac{1}{3} \unab (\phi_0 (\ux ,t) +2\phi_2 (\ux ,t)  ) +\sigt(\ux ,t)\ul{\phi}_0 (\ux ,t)  = 0, \\
\recv \D{\ul{\phi}_3}{t}(\ux ,t) &+ \unab \left ( \frac{3}{7} \phi_2 (\ux ,t)  \right) +\sigt(\ux ,t)\ul{\phi}_3 (\ux ,t)  = 0,
}
where the three-dimensional vectors $\ul{\phi}_1(\ux, t)$ and $\ul{\phi}_3(\ux, t)$ are obtained by formally replacing the scalars ${\phi}_1 (x, t)$ and $\phi_3 (x, t)$ in \refeqs{clDer:eq8} by vectors.

Although \refeqs{clDer:eq18} differ from \refeqs{clDer:eq15} it is
possible to transform them into each other by a similarity transformation acting on the PDE system of \refeqs{clDer:eq15}.
This transformation is defined by
\balign{
\left [\phi_0,\phi_2, \ul{J}_0,\ul{J}_2 \right ] \mapsto \left [\phi_0,\phi_2, \ul{\phi}_1, \frac25 \ul{\phi}_1 + \frac35 \ul{\phi}_3 \right].
}
Consequently, \refeqs{clDer:eq15} and \refeqs{clDer:eq18} are equivalent and their solutions are either identical or can easily be transformed into one another.
\end{rem}

The above remark confirms that numerical codes, based on time-dependent $SP_3$ equations obtained by simply adding the time-derivative, in fact do approximate equations which are equivalent to those derived in this paper. 
However, since all existing codes are based on equations derived in a purely heuristic way, we now wish to provide a theoretical foundation for these equations, which are already being solved numerically.

\section{Formal Asymptotic Derivation}
\label{sec:asympDer}

To keep our discussion simple, we restrict the asymptotic analysis to the monoenergetic, \, 3-D isotropically scattering particle transport in a \textit{homogeneous} medium.
However, a similar analysis might also be performed for anisotropic scattering, which is already presented in \cite{Lar11a} for the steady-state equations.
As in Section \ref{sec:clDer}, we start our analysis with the linear Boltzmann equation:
\seqalign{\label{asympDer:eq1}}{
\recv \D{\Psi}{t} (\ux ,\ul{\Omega} ,t) +\ul{\Omega} \cdot\unab \psix +\sigt\psix &= \frac{1}{4\pi} ( \sigs \phi_0(\ux ,t) +Q(\ux ,t)), \label{asympDer:eq1a}
\intertext{or, dividing by $\sigt$}
{\mathcal T} \psix + \ul{\Omega}\cdot \ul{\calX} \psix +\psix &= \frac{1}{4\pi} \left ( c\phi_0(\ux ,t) +\frac{Q(\ux ,t)}{\sigt} \right ), \label{asympDer:eq1b}
}
where
\seqalign{\label{asympDer:eq2}}{
\phix &= \int_{S^2} \psix d\Omega\, \equiv\, \tx{scalar flux}, \\
{\mathcal T} &= \frac{1}{v\sigt} \D{}{t}, \\
\ul{{\mathcal X}} &=\frac{1}{\sigt} \unab,\\
c &= \frac{\sigs}{\sigt}\,  \equiv\, \tx{scattering ratio}.
}

%
%
%
\noindent If we define the angular projection operator
\seqalign{\label{asympDer:eq3}}{
({\mathcal P}\psi ) (\ux ,t) &:= \frac{1}{4\pi} \int_{S^2} \psix d\Omega,
\intertext{then}
({\mathcal P}\psi) (\ux ,t) &= \frac{1}{4\pi} \phix \quand  ({\mathcal P}\ul{\Omega}\psi  )(\ux ,t) = \frac{1}{4\pi} \ul{J}(\ux ,t),
\intertext{where}
\ul{J}(\ux ,t) &= \int_{S^2} \ul{\Omega}\psix d\Omega \, \equiv \, \tx{current}. \label{asympDer:eq7}
}
Operating on \refeq{asympDer:eq1b} by $4\pi {\mathcal P}$, we obtain
\begin{align}
{\mathcal T} \phi_0(\ux ,t) + \ul{\calX}\cdot \ul{J}(\ux ,t)  + \phix = \left ( c\phix + \frac{Q(\ux ,t)}{\sigt}  \right ). \label{asympDer:eq8}
\end{align}
We also operate on \refeq{asympDer:eq1b} by $(I -{\mathcal P})$ to get the additional equation
\begin{align}
(I -{\mathcal P}) {\mathcal T} \psix + (I -{\mathcal P}) \ul{\Omega}\cdot \ul{{\mathcal X}} \psix + (I -{\mathcal P})\psix &= 0,
\end{align}
or
\seqalign{\label{asympDer:eq4}}{
(I +{\mathcal T}) \psix + (I -{\mathcal P}) \ul{\Omega}\cdot \ul{ {\mathcal X}} \psix &= {\mathcal P} (I +{\mathcal T})\psix  \\
&=  (I +{\mathcal T}) {\mathcal P} \psix \\
&= \frac{1}{4\pi}  (I +{\mathcal T}) \phix.
}
This implies
\begin{align}
\psix + \left [ (I +{\mathcal T})^{-1} (I-{\mathcal P}) \ul{\Omega}\cdot\ul{{\mathcal X}}  \right ] \psix = \frac{1}{4\pi} \phix. \label{asympDer:eq5}
\end{align}
If we define the operator ${\mathcal M}$ by
\begin{align}
{\mathcal M} &:= \invOp (I- {\mathcal P}), \label{asympDer:eq9}
\intertext{then \refeq{asympDer:eq5} can be written as}
\psix &+ \momegax\psix = \frac{1}{4\pi} \phix,
\intertext{or}
\psix &=  \frac{1}{4\pi} (I + \momegax)^{-1} \phix. \label{asympDer:eq6}
\end{align}
Introducing \refeq{asympDer:eq6} into the definition of the current in \refeq{asympDer:eq7}:

\begin{align}
\ul{J} (\ux ,t) &= \int_{S^2} \ul{\Omega}\psix d\Omega \\
&= \frac{1}{4\pi} \int_{S^2} \ul{\Omega}  ( I+\momegax )^{-1} \phix d\Omega,
\end{align}
and then introducing this result into \refeq{asympDer:eq8}, we end up with:
\begin{align}
\recv \D{\phi_0}{t} (\ux ,t) +\frac{1}{4\pi} \int_{S^2} \ul{\Omega}\cdot \unab  &( I+\momegax )^{-1} \phix d\Omega+\siga\phix = Q(\ux ,t). \label{asympDer:eq16}
\end{align}
This equation for $\phi_0(\ux ,t)$ is formally exact, but the integral term is a complicated operator which needs further simplifications.
We introduce a small, positive, dimensionless parameter $\eps$ into \refeq{asympDer:eq16} such that the operator $\ds \frac{1}{\sigt} \unab$ becomes small, i.e.,
\balign{
\ul{{\mathcal X}} &= \eps \frac{1}{\sigt} \unab \label{asympDer:eq2c}.
}
\begin{rem}
It should be emphasized that the only assumption for the following asymptotic analysis is that $\ul{\calX}=\order{}$.
Neither the time derivative is supposed to be small nor source terms are scaled.
This is purely formal and is chosen to keep the framework as general as possible.
To draw a line from the scaling considered in this paper to scalings from previous asymptotic $SP_N$ derivations in literature, we list some of them which are acceptable for our asymptotics:
\begin{itemize}
\item \textbf{conventional scaling} \cite{HabMat75, Lar80a, LarKel74, LarMorMcG93, Pom93}: Here, the system is assumed to be scattering-dominated with its collision rate being much larger than its absorption rate. In this case, $\eps$ is the ratio of the mean free path (which corresponds to $\sigt^{-1}$) over a typical length scale for the solution. This scaling is the standard scaling, which has been used to gain the diffusion or \textit{steady-state} $SP_N$ equations by performing an asymptotic analysis for $\eps\approx 0$.

\item \textbf{generalized conventional scaling} \cite{Lar11a}: \textit{Larsen} introduces an alternate scaling which is physically consistent with the conventional scaling and additionally, includes free parameters. Depending on the choice of these parameters either the standard or modified diffusion and $SP_N$ equations are obtained. From the theoretical point of view, the latter equations are proven to increase the accuracy for deep penetration problems.

\item \textbf{parabolic scaling} \cite{FraKlaLar07}: An asymptotic approach for the time-dependent $SP_3$ equations is presented in which the time derivative, source term and absorption cross section are scaled by $\eps^2$ and the space derivative by $\eps$.
This scaling can also be achieved by introducing
\seqalign{}{
 v &= \frac{ \tilde{v} }{\eps} \\
\sigt &= \frac{\sigma_t}{\eps}, \\
\siga &= \eps\sigma_a , \\
Q(\ux ,t) &= \eps q(\ux ,t),
}
where $\tilde{v}, \sigma_t, \sigma_a, q$ are of ${\mathcal O}(1)$, into \refeq{intro:eq1} and dividing by $\sigt$.
In addition to the physical assertions from the conventional scaling, it also requires that particles travel at high velocities.
Combined with a high collision rate, low absorption rate, and small source terms, this scaling as a whole implies a slowly varying solution in space and  an even smaller variation in time.
\end{itemize}

\end{rem}
\noindent Next, we asymptotically expand the operator
\balign{
{\mathcal L} = \frac{1}{4\pi} \int_{S^2}  \ul{\Omega}\cdot \unab  ( I+\momegax )^{-1} d\Omega. \label{asympDer:eq11}
}
Due to the assertion in \refeq{asympDer:eq2c} about the dimensionless spatial gradient $\ul{{\mathcal X}} = \eps \frac{1}{\sigt} \unab$
or any scaling which yields $\ul{\mathcal X} = \order{}$, we can expand the operator ${\mathcal L}$ in \refeq{asympDer:eq11} in a Neumann series. \\ \\
Thus, for $\eps$ sufficiently small, we have that
\balign{
{\mathcal L} &= \sum_{n=0}^{\infty} (-1)^n {\mathcal L}_n,  \label{asympDer:eq13}
\intertext{where}
{\mathcal L}_n &= \left [ \frac{1}{4\pi} \int_{S^2} \uomega\cdot\unab ( \momegax )^n  \right ] = {\mathcal O} (\eps^n).
}
To achieve an ${\mathcal O} (\eps^7)$ approximation we need the first seven ${\mathcal L}_n$:
\seqalign{\label{asympDer:eq12}}{
{\mathcal L}_0 &= 0, \\
{\mathcal L}_1 &= \frac{1}{3} \unab\cdot \invOp\ul{\calX}  \\
{\mathcal L}_2 &= 0,  \\
{\mathcal L}_3 &= \frac{4}{45} \left [\unab\cdot \invOp\ul{\calX} \right ] \invOp \left [ \ul{\calX}\cdot\invOp\ul{\calX}  \right ] \label{asympDer:eq12d} \\
{\mathcal L}_4 &= 0,  \\
{\mathcal L}_5 &= \frac{44}{945} \left [\unab\cdot \invOp\ul{\calX} \right ] \invOp \left [ \ul{\calX}\cdot\invOp\ul{\calX}  \right ]  \invOp \left [ \ul{\calX}\cdot\invOp\ul{\calX}  \right ] \label{asympDer:eq12f} \\
{\mathcal L}_6 &= 0,
}
which are calculated in detail in the Appendix.

\begin{rem}
It is important to emphasize one crucial assumption made in the derivation of \refeq{asympDer:eq12d} and \refeq{asympDer:eq12f}: Both are only exact for either homogeneous media or a system in which $\sigt$ depends only on one spatial variable.
However, the rest of \refeqs{asympDer:eq12} are also exact for heterogeneous media.
\end{rem}

\subsection{$SP_1$ Equations}

Ignoring terms of $\order{3}$ in \refeq{asympDer:eq13} we obtain from \refeqs{asympDer:eq12}
\[ {\mathcal L} = - \frac{1}{3} \unab\cdot \invOp\ul{\calX} +\order{3}. \]
Introducing this approximation for ${\mathcal L}$ in \refeq{asympDer:eq11} and \refeq{asympDer:eq16} we get
\balign{
\recv \D{\phi_0}{t} (\ux ,t) - \frac{1}{3} \unab\cdot \invOp\ul{\calX} \phix +\siga\phix +\order{3} = Q(\ux ,t).
}
If we additionally define
\[ \ul{J}_0 (\ux ,t) := -\frac13 \invOp\ul{\calX} \phix \]
and drop the error term, the above equations simplify to the $SP_1$ equations:
\seqalign{\label{asympDer:SP1}}{
\recv \D{\phi_0}{t} (\ux ,t) &+ \unab\cdot \ul{J}_0 (\ux ,t) +\siga(\ux, t) \phix = Q(\ux ,t), \\
\recv \D{\ul{J}_0}{t} (\ux ,t) &+ \frac13 \unab \phix +\sigt(\ux ,t) \ul{J}_0 = 0.
}

\begin{rem}
Whereas the steady-state $SP_1$ approximation is the standard diffusion equation, which requires only one scalar-valued function $\phi_0$, \refeqs{asympDer:SP1} are a system of two equations with the scalar variable $\phi_0$ and vector $\ul{J}_0$.
By simply adding the time-derivative to the steady-state $SP_1$ equation one might expect a parabolic time-dependent $SP_1$ approximation which is also obtained in \cite{FraKlaLar07}. 
However, above time-dependent $SP_1$ equations are hyperbolic.
\end{rem}

\subsection{$SP_2$ Equations}
More accurate solutions can be gained systematically by taking higher order terms in \refeq{asympDer:eq13} into account.
An asymptotically higher order approximation of $\order{5}$ is given by:
\balign{
\recv \D{\phi_0}{t} (\ux ,t) - \frac{1}{3} \sigt \, L \left [  I + \frac{4}{15} \invOp L \right ] \phix + \siga\phix  +\order{5} = Q(\ux ,t).	\label{asympDer:eq19}
}
where, for the sake of notational simplicity, the operator $L$ is defined as
\[ L :=  \left [ \ul{\calX}\cdot\invOp\ul{\calX}  \right ]  = \order{2}. \]
We approximate the operator in square brackets in \refeq{asympDer:eq19} by
\balign{
\left [  I + \frac{4}{15} \invOp L \right ] = \left [  I - \frac{4}{15} \invOp L \right ]^{-1} +\order{4}, \label{asympDer:eq21}}
and set
\balign{2\phi_2 (\ux ,t) +\phix :=  \left [  I - \frac{4}{15} \invOp L \right ]^{-1} \phix, }
which can be rewritten to
\balign{\frac{4}{15} \invOp L ( 2\phi_2 (\ux ,t)  +\phix )&= 2\phi_2 (\ux ,t).
\intertext{Hence, by applying the operator $(I+{\mathcal T})$ on the left, we get}
\recv\D{}{t} \phi_2(\ux ,t) + \sigt \phi_2(\ux ,t) &= \frac{2}{15} \unab\cdot\invOp\ul{\calX} (\phix +2\phi_2 (\ux ,t)). \label{asympDer:eq20}
}
Combining \refeq{asympDer:eq19} and \refeq{asympDer:eq20} as well as discarding the error term, we obtain the system
\seqalign{}{
\recv \D{\phi_0}{t} (\ux ,t) &+ \unab\cdot \ul{J}_0 (\ux ,t) +\siga( \ux ,t) \phix = Q(\ux ,t),\\
\recv \D{\phi_2}{t} (\ux ,t) &+ \unab\cdot \ul{J}_2 (\ux ,t) +\sigt (\ux ,t)\phi_2(\ux, t) = 0, \\
\recv \D{\ul{J}_0}{t} (\ux ,t) &+ \frac13 \unab \bigg( \phix +2\phi_2(\ux, t) \bigg)  +\sigt(\ux ,t) \ul{J}_0 (\ux ,t) = 0, \\
\recv \D{\ul{J}_2}{t} (\ux ,t) &+ \frac{2}{15} \unab \bigg( \phix +2\phi_2(\ux, t) \bigg)  +\sigt(\ux ,t) \ul{J}_2 (\ux ,t) = 0.
}

%
%
\subsection{$SP_3$ Equations}
Having collected all operators, we approximate ${\mathcal L}$ by truncating the series in \refeq{asympDer:eq13} at $n=6$ and introducing \refeqs{asympDer:eq12} into \refeq{asympDer:eq13}:
\seqalign{\label{asympDer:eq14}}{
{\mathcal L} = &- \frac{1}{3} \unab\cdot \invOp\ul{\calX} \\
& - \frac{4}{45} \left [\unab\cdot \invOp\ul{\calX} \right ] \invOp \left [ \ul{\calX}\cdot\invOp\ul{\calX}  \right ] \\
& -  \frac{44}{945} \left [\unab\cdot \invOp\ul{\calX} \right ] \left ( \invOp \left [ \ul{\calX}\cdot\invOp\ul{\calX}  \right ]  \right )^2+\order{7} \\
= & - \frac{1}{3} \unab\cdot \invOp\ul{\calX} \left \{  I + \frac{4}{15} \invOp \left [ \ul{\calX}\cdot\invOp\ul{\calX}  \right ] \right . \nonumber \\
& \hspace{5cm} \left . + \frac{44}{315} \left (  \invOp \left [ \ul{\calX}\cdot\invOp\ul{\calX}  \right ] \right )^2 \right \} +\order{7}. \label{asympDer:eq14d}
}
We rewrite \refeq{asympDer:eq14d} to
\seqalign{}{
{\mathcal L} = & - \frac{1}{3} \sigt \, L \left \{  I + \frac{4}{15} \invOp L + \frac{44}{315} \left (  \invOp L \right )^2 \right \} +\order{7} \\
= & - \frac{1}{3} \sigt \, L \left \{  I + \left [ I + \frac{11}{21} \invOp L \right ] \frac{4}{15} \invOp L \right \} +\order{7}. \label{asympDer:eq15b}
}
Approximating the term in square brackets of \refeq{asympDer:eq15b} like in \refeq{asympDer:eq21}, we conclude
\balign{
{\mathcal L} = - \frac{1}{3} \sigt \, L \left \{  I + \left [ I - \frac{11}{21} \invOp L \right ]^{-1} \frac{4}{15} \invOp L \right \} +\order{7}.
}
Using this approximation for ${\mathcal L}$ in \refeq{asympDer:eq11} and \refeq{asympDer:eq16}, we get:
\balign{
\recv &\D{\phi_0}{t} (\ux ,t) + \siga\phix = Q(\ux ,t) \nonumber \\
&+ \frac{1}{3} \sigt \, L \left \{  \phix + \left [ I - \frac{11}{21} \invOp L \right ]^{-1} \frac{4}{15} \invOp L\phix \right \} +\order{7}.
}
If we define
\balign{
2\phi_2(\ux ,t) := \left [ I - \frac{11}{21} \invOp L \right ]^{-1} \frac{4}{15} \invOp L\phix,
}
then $\phi_2$ satisfies
\balign{
 \left [ I - \frac{11}{21} \invOp L \right ] \phi_2(\ux ,t) = \frac{2}{15} \invOp L\phix.
}
Operating by $(I +{\mathcal T})$ on the last equation yields
\balign{
\left [ I +{\mathcal T} - \frac{11}{21} L \right ] \phi_2(\ux ,t) &= \frac{2}{15} L\phix,
}
or
\balign{
\left ( {\mathcal T} + I \right ) \phi_2(\ux ,t) &= L \left ( \frac{2}{15} \phix + \frac{11}{21}\phi_2(\ux , t)  \right ) = \frac{1}{3} L \left ( \frac{2}{5} \phix +\frac{11}{7}\phi_2(\ux , t) \right ),
}
which is equivalent to
\balign{
\left ( \recv\D{}{t} + \sigt \right ) \phi_2(\ux ,t) = \frac{1}{3} \unab \cdot \invOp \frac{1}{\sigt} \unab \left ( \frac{2}{5} \phix +\frac{11}{7}\phi_2(\ux , t)  \right ).
}
Altogether, we get
\seqalign{\label{asympDer:eq17}}{
\recv\D{\phi_0}{t} (\ux ,t)  + \siga \phix &= \unab \cdot \invOp \frac{1}{3\sigt} \unab \left (\phix +2\phi_2(\ux , t)  \right ) + Q(\ux ,t), \\
\recv\D{\phi_2}{t} (\ux ,t) + \sigt \phi_2(\ux ,t) &= \unab \cdot \invOp \frac{1}{3\sigt} \unab \left ( \frac{2}{5} \phix +\frac{11}{7}\phi_2(\ux , t)  \right ).
}
In a homogeneous medium, eqs.~(\ref{asympDer:eq17}) are identical to \refeqs{clDer:eq13}.
This essentially proves the asymptotic derivation of the previously obtained \refeqs{clDer:eq13}. 
One can continue rewriting these equations to \refeqs{clDer:eq15} in the same way as is done in Section \ref{sec:clDer}.

\section{Numerical Results in 2D}
\label{sec:numres}

We perform 2-D simulations for diffusion, the $P_N$, and the $SP_N$ equations. The computations for the latter two approximations are done with a version of the code \texttt{StaRMAp} by Seibold and Frank \cite{SeiboldFrank2012}. The name \texttt{StaRMAp} stands for ``\texttt{Sta}ggered grid \texttt{R}adiation \texttt{M}oment \texttt{Ap}proximation'',
which describes the key methodology of the approach. 
More specifically, it is a second order accurate finite difference method for linear hyperbolic balance laws of the form
\begin{equation}
\label{eq:hyperbolic_balance_law}
\partial_t\vec{u} + M_x\cdot\partial_x\vec{u} + M_y\cdot\partial_y\vec{u}
+C\cdot\vec{u} = \vec{q}\;,
\end{equation}
where the matrices $M_x$, $M_y$, and $C$ possess specific patterns of their nonzero entries, as described below. The numerical method is implemented in a concise \textsc{Matlab} code that the authors plan to make publicly available upon submission of the corresponding paper \cite{SeiboldFrank2012}. Let the components of the solution vector $\vec{u}$ be indexed by $\{1,2,\dots,S\}$. The requirement on the nonzero entry patterns of $M_x$, $M_y$, and $C$ is that the components of $\vec{u}$ can be distributed into four disjoint sets, according to $\{1,2,\dots,S\} = I_{00} \,\dot{\cup}\, I_{10} \,\dot{\cup}\, I_{01} \,\dot{\cup}\, I_{11}$, such that the following properties hold:
\begin{equation}
\label{eq:matrix_conditions}
\begin{split}
(M_x)_{i,j} = 0\; &\forall\,(i,j)\notin ((I_{00}\times I_{10})
\cup (I_{10}\times I_{00}) \cup (I_{01}\times I_{11}) \cup (I_{11}\times I_{01}))\;, \\
(M_y)_{i,j} = 0\; &\forall\,(i,j)\notin ((I_{00}\times I_{01})
\cup (I_{01}\times I_{00}) \cup (I_{10}\times I_{11}) \cup (I_{11}\times I_{10}))\;, \\
C_{i,j} = 0\; &\forall\,(i,j)\notin ((I_{00}\times I_{00})
\cup (I_{10}\times I_{10}) \cup (I_{01}\times I_{01}) \cup (I_{11}\times I_{11}))\;.
\end{split}
\end{equation}
With this distribution of the indices of the solution components, we consider the following four fully staggered sub-grids
\begin{align*}
G_{00} = \{(i\Delta x,j\Delta y) \,|\, i,j\in\mathbb{Z}\}\;,
& \quad G_{10} = \{((i+\tfrac{1}{2})\Delta x,j\Delta y) \,|\, i,j\in\mathbb{Z}\}\;, \\
G_{01} = \{(i\Delta x,(j+\tfrac{1}{2})\Delta y) \,|\, i,j\in\mathbb{Z}\}\;,
& \quad G_{11} = \{((i+\tfrac{1}{2})\Delta x,(j+\tfrac{1}{2})\Delta y) \,|\, i,j\in\mathbb{Z}\}\;,
\end{align*}
and assign the components with indices in $I_{k\ell}$ to the corresponding sub-grid $G_{k\ell}$, where $k,\ell\in\{0,1\}$. On these fully staggered grids, any spatial derivative is approximated by a simple central difference stencil:
\begin{align*}
\partial_x w(i\Delta x,j\Delta y) &\approx \tfrac{1}{\Delta x}
\left(w((i+\tfrac{1}{2})\Delta x,j\Delta y)-w((i-\tfrac{1}{2})\Delta x,j\Delta y)\right)
\;\forall\,i,j\in\tfrac{1}{2}\mathbb{Z}\;, \\
\partial_y w(i\Delta x,j\Delta y) &\approx \tfrac{1}{\Delta y}
\left(w(i\Delta x,(j+\tfrac{1}{2})\Delta y)-w(i\Delta x,(j-\tfrac{1}{2})\Delta y)\right)
\;\forall\,i,j\in\tfrac{1}{2}\mathbb{Z}\;.
\end{align*}
Hence, $x$-derivatives of components on $G_{k\ell}$ live on $G_{1-k,\ell}$, and $y$-derivatives of components on $G_{k\ell}$ live on $G_{k,1-\ell}$, where $k,\ell\in\{0,1\}$. The nonzero entry patterns \eqref{eq:matrix_conditions} guarantee that the distribution of the indices of $\vec{u}$ into the sets $I_{00}$, $I_{10}$, $I_{01}$, and $I_{11}$ is identical to the corresponding distribution of the indices of $M_x\cdot\partial_x\vec{u} + M_y\cdot\partial_y\vec{u}+C\cdot\vec{u}$. Both the classical $P_N$ equations as well as the here derived $SP_N$ equations, possess precisely the nonzero entry patterns \eqref{eq:matrix_conditions}, that admit a solution on fully staggered grids.

The time-derivative in \eqref{eq:hyperbolic_balance_law} is resolved by bootstrapping. Having a time step $\Delta t$, we associate the components that live on $G_{00}\cup G_{11}$ with the times $T_0 = \{n\Delta t \,|\, n\in\mathbb{Z}\}$, and the components that live on $G_{10}\cup G_{01}$ with the times $T_1 = \{(n+\tfrac{1}{2})\Delta t \,|\, n\in\mathbb{Z}\}$. A full time step consists of two sub-steps: first, update information on the grid $G_{10}\cup G_{01}$ from time $(n-\tfrac{1}{2})\Delta t$ to $(n+\tfrac{1}{2})\Delta t$, where information on $G_{00}\cup G_{11}$ at the mid-time $n\Delta t$ is used; second, update information on the grid $G_{00}\cup G_{11}$ from time $n\Delta t$ to $(n+1)\Delta t$, where information on $G_{10}\cup G_{01}$ at the mid-time $(n+\tfrac{1}{2})\Delta t$ is used.

Specifically, the sub-step update rule is implemented as follows (here for $G_{00}\cup G_{11}$; the other sub-step works analogously). The terms $\vec{r} = \vec{q}-M_x\cdot\partial_x\vec{u}-M_y\cdot\partial_y\vec{u}$ that come from $G_{10}\cup G_{01}$ at the mid-step time are considered constant over the sub-step. Thus, equation \eqref{eq:hyperbolic_balance_law} becomes the ODE
\begin{equation}
\label{eq:sub_step_evolution}
\partial_t\vec{u} + C\cdot\vec{u} = \vec{r}
\end{equation}
with $\vec{r} = \text{const}$. In the special case that $C$ is a diagonal matrix, $C = \text{diag}(c_1,\dots,c_S)$, we can solve \eqref{eq:sub_step_evolution} from $n\Delta t$ to $(n+1)\Delta t$ explicitly:
\begin{equation}
\label{eq:sub_step_solution}
u_k(x,(n+1)\Delta t) = \exp(-c_k(x)\Delta t)u_k(x,n\Delta t)
-\tfrac{1}{c_k}(1-\exp(-c_k(x)\Delta t))r_k\;.
\end{equation}
Note that, if $C$ is time-dependent, we evaluate it at the mid-step time, in which case \eqref{eq:sub_step_solution} becomes a second-order accurate approximation to the true solution of \eqref{eq:sub_step_evolution}.
This second-order convergence was verified by the method of manufactured solutions \cite{SalKnu00} for the $SP_N$ equations of order $N=1,3$.


As $P_N$ as well as $SP_N$ solutions are spatially discontinuous at material interfaces for even $N$ \cite{TomLar96}, we only present numerical calculations for \textit{odd} N in the following.
In some applications the diffusion equation is used to calculate approximations to the Boltzmann equation.
Concerning computational effort, this approach is comparably inexpensive for coarse to moderate spatial resolutions, in which case the resulting linear systems can be solved rapidly. 
Unfortunately, its accuracy is not satisfactory, as will be shown for example in Section~\ref{subsec:checkerboard}.
For the sake of comparison, we solve the following time-dependent diffusion equation
\balign{
\label{numres:eq1}
\recv \D{\phi_0}{t} (x,y,t) &= \unab\cdot \left [  D(x,y,t) \unab \phi_0(x,y,t) \right ] - \siga (x,y,t)\phi_0(x,y,t) + Q(x,y,t), \\
D(x,y,t) &= \frac{1}{3\sigt(x,y,t)},
}
which can be interpreted as a low-order approximation to the Boltzmann equation (\ref{intro:eq1}) \cite{FraKlaLar07}.
We apply the second order central in space -- Crank-Nicolson in time Finite Difference scheme to discretize above equation.

%
%

\subsection{Equivalence of $SP_N$ and $P_N$ equations}
\label{sec:PnSpnEquiv}

The detailed analysis of the
(simplified) spherical harmonics approximation brought along certain conditions under which the steady-state $P_N$ and $SP_N$ equations are equivalent.
McClarren describes some of them and demonstrates numerically
this equivalence on a square of dimension $L=5$ in a homogeneous medium with isotropic material coefficients and sources \cite{McC11}.
We slightly modify the inhomogeneous source from \cite{McC11} and change it to the following \textit{time-dependent} sinusoidal term
\balign{
Q(x,y,t) = \left ( 2+\sin(4\pi t)e^{-t/3} \right )\cdot	\begin{cases}
			  																	1,      & (x,y) \in [1.75, 2.25]\times [1.75, 2.25], \\
			  																	1, 	   & (x,y) \in [2.75, 3.25]\times [1.5, 2.5], \\
			  																	1,      & (x,y) \in [1.75, 2.25]\times [2.75, 3.25], \\
			  																	1,      & (x,y) \in [3.5, 4.25]\times [3.5, 3.75], \\
			  																	0,	   & \text{ otherwise}.
			 								 									\end{cases}
}
All criteria for the equivalence were developed for the \textit{steady-state} equations, and there is no obvious reason why they should also be true for the time-dependent case.
We implement this test case with $\sigt=1$ and $\siga=0.9$ and investigate the temporal as well as steady-state behavior of various approximations.
Periodic boundary conditions are enforced for all methods.

This problem has an absorption coefficient which is nine times larger than the scattering coefficient.
Due to the inhomogeneous source, the solution has additionally large spatial gradients.
Consequently, an agreement between $P_N$ and $SP_N$ solutions cannot be justified by any of the asymptotic scalings mentioned in Section \ref{sec:asympDer}.

\begin{figure}[h!]
 \centering
 \subfloat[$SP_9$]{\includegraphics[scale=0.7]{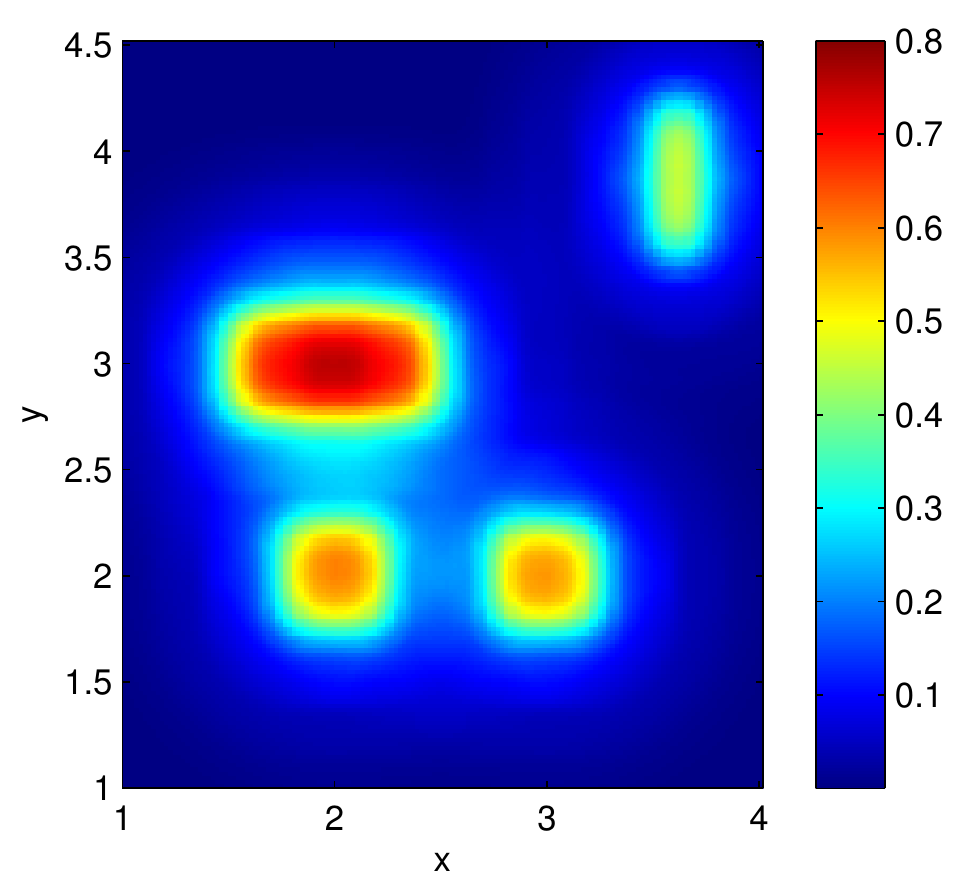}}
  \,
 \subfloat[$P_{9}$]{\includegraphics[scale=0.7]{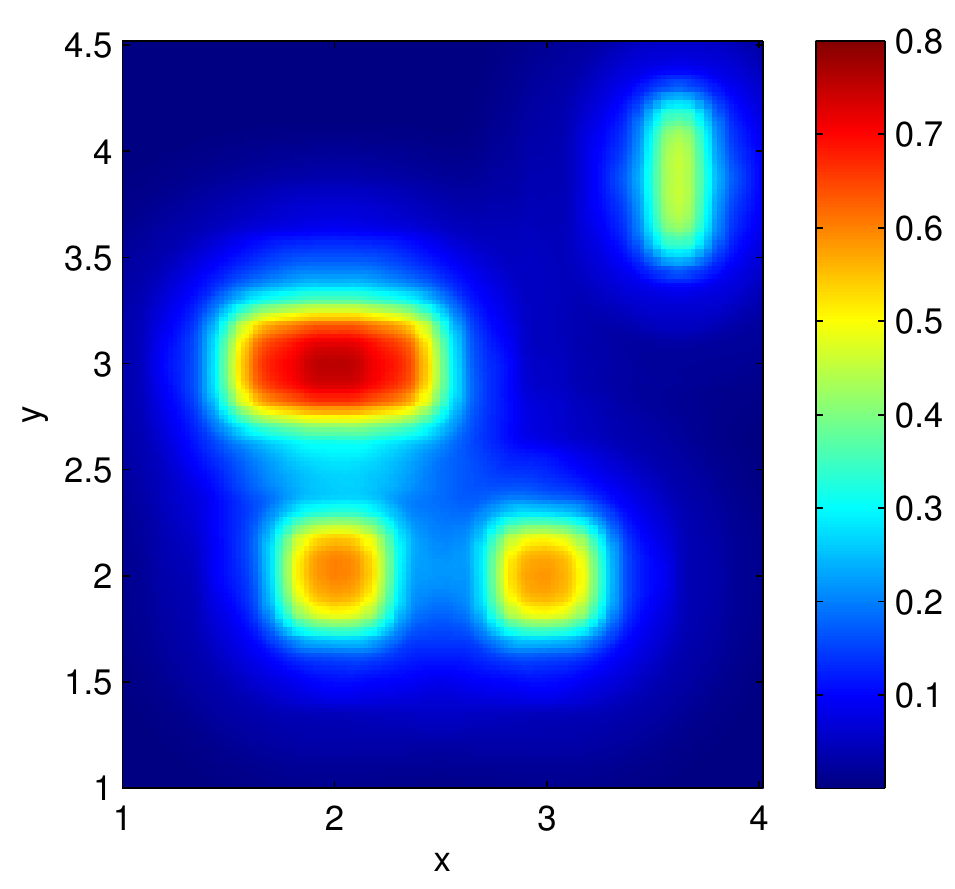}}
 \\
\subfloat[Diffusion]{\includegraphics[scale=0.7]{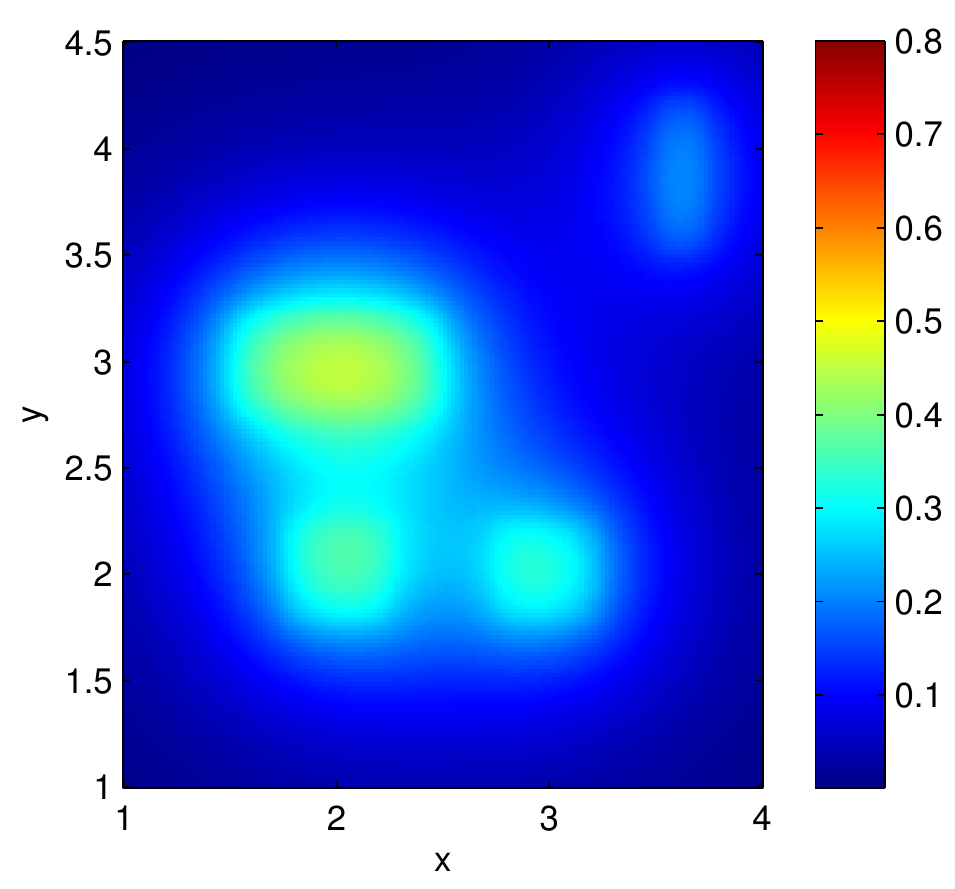}}
  \,
 \subfloat[$P_{39}$]{\includegraphics[scale=0.7]{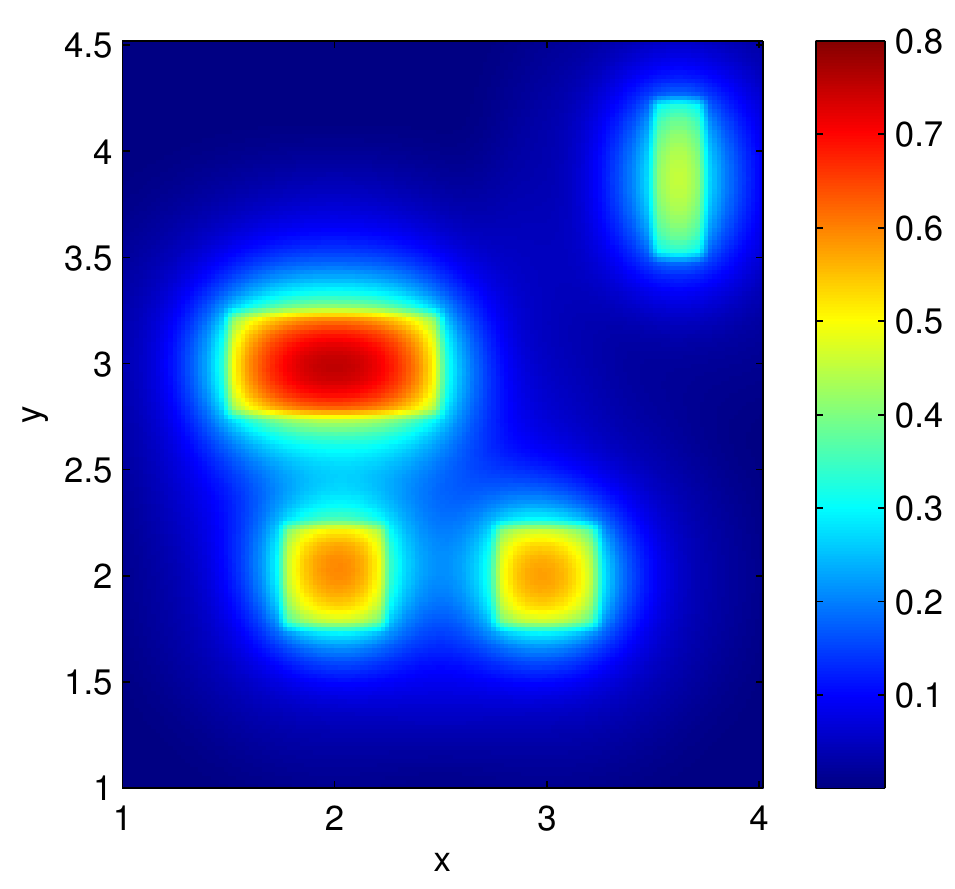}}
\caption{Approximations to $\phi_0$ at $t_{\tx{final}}=1$, $250\times250$ discretization points.}
  \label{pict:Equiv2d}
 \end{figure}

\begin{figure}
 \centering
 \includegraphics[scale=1]{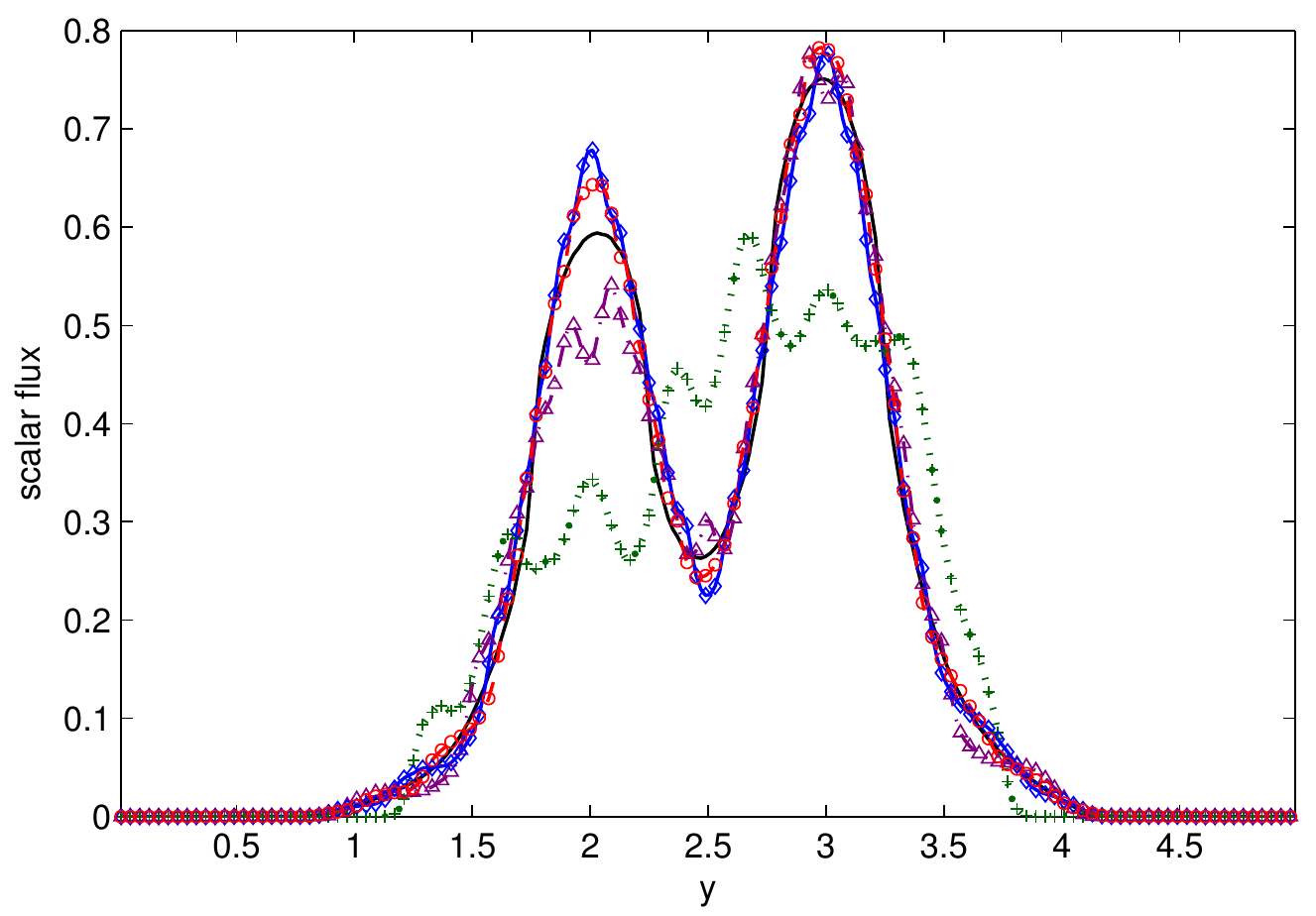}
\caption{$\phi_0$ at $t_{\tx{final}}=1$ along $x=2$, $250\times250$ discretization points: $P_1$ (green dotted line), $SP_1$ (green crosses), $P_3$ (purple dash-dot line), $SP_3$ (purple triangles), $P_5$ (blue solid line), $SP_5$ (blue diamonds), $P_9$ (red dashed line), $SP_9$ (red circles), $P_{39}$ (black solid line)}
  \label{pict:xcut}
 \end{figure}

\reffig{pict:Equiv2d} displays the scalar flux $\phi_0$ at $t=1$ from several methods.
Note that $\phi_0$ is not in steady-state at this time.
As the condition of small spatial derivatives, which underlie the diffusion approximation, is not met in this situation, the diffusion solution is inaccurate.
The $SP_9$ and $P_9$ solutions appear to be identical, which is also confirmed in \reffig{pict:xcut} showing the scalar flux along $x=2$ for (simplified) spherical harmonics of order $N=1,3,5,9$.
To the eye, there is again no difference between $SP_N$ and $P_N$.
Moreover, we observe a significant improvement from the $SP_1$ to the $SP_3$ approximation.
Although $SP_5$ still shows significant deviations, an order of $N=9$ is sufficient for the simplified $P_N$ approximation to be very close to the high-order $P_{39}$ solution, which is considered as reference here.

It turns out that all $SP_N$ and $P_N$ solutions are indeed equivalent.
To verify this behavior 
more precisely we calculate the difference in the $L^{\infty}$-norm
\[ \max_{i,j}{|\phi_0^{SP_N}(x_i,y_j)-\phi_0^{P_N}(x_i,y_j)|}\]
at several times for $N=1,3,5,7,11$ and observe that all differences are as small as the machine precision.
Hence, not only do the $P_N$ and $SP_N$ equations have the same solutions but also the numerical discretizations of them, which is an advantageous property of the applied numerical scheme.  
Our numerical experiments indicate that the analysis, performed for the steady-state equations, to prove the equivalence of $P_N$ and $SP_N$ equations in a general, homogeneous medium with isotropic cross sections and sources might be extended to the time-dependent equations including time-dependent sources.
This is an issue of future work which goes beyond the purpose of this paper.

Note that in this case of a 2-D problem the $P_N$ equations have $\frac{(N+1)(N+2)}{2}$ unknowns, whereas the $SP_N$ method consists only of $3(\lfloor \frac{N}{2} \rfloor +1 )$ unknowns.
Nevertheless, it is remarkable that the $P_N$ and $SP_N$ solutions in this example agree even for the time-dependent case.
Hence, there exist situations where high-order $SP_N$ approximations yield very accurate solutions (here for $N=9$), and the diffusion approximation gives unsatisfactory results.

\subsection{A Moving Rod}

The prediction of the behavior of nuclear reactors is essential for the design and safe operation of nuclear power plants.
Apart from many other interactions, one challenge is to develop efficient and accurate techniques for the description of neutron distributions.

We consider the linear transport Boltzmann equation (\ref{intro:eq1}) with slight modifications as before.
Neglecting energy dependence and the coupling to precursors, approximative models to the following equation on $[-\frac{L}{2},\frac{L}{2}]\times [-\frac{L}{2},\frac{L}{2}]$ are solved:
\balign{
\frac{1}{v} \D{\Psi}{t} (x,y,\ul{\Omega},t) +\ul{\Omega}\cdot \ul{\nabla} \Psi(x,y,\ul{\Omega},t) &+ [\sigs +\Sigma_f +\Sigma_{\gamma}(x,y,t) ] \Psi(x,y,\ul{\Omega},t) \nonumber \\
 &= \frac{\Sigma_s +\nu\Sigma_f}{4\pi} \int_{S^2} \Psi(x,y, \ul{\Omega}',t) d\Omega' . \label{numres:eq3}
}

In particular, the initial condition for the scalar flux is set to $\phi_0(x,y,0)=10^6$ and all other variables to zero.
Periodic boundary conditions are enforced, and we use the following parameters:
\balign{
v = 1, \q L =2, \q \nu = 0.9, \q \Sigma_f = 2, \q \Sigma_s = 1.
}

In addition to the scattering process (taken into account by $\sigs$), the above equation includes two relatively frequent interactions: $\Sigma_f$ is the fission cross section describing the probability that a neutron will initiate a fission event when it collides with a nucleus;
$\Sigma_{\gamma}(x,y,t)$ is the capture cross section characterizing a capture event in which the nucleus gains a neutron.
Consequently, the absorption cross section is the sum of both:
\[\siga(x,y,t) = \Sigma_f  +\Sigma_{\gamma}(x,y,t). \]
In a fission event, the target nucleus splits into two daughter nuclei and $\nu$ is usually the mean number of fission neutrons that are released.
However, to keep the test case simpler, we use $0\leq \nu\leq 1$ as a parameter and set the capture cross section to
\balign{
\Sigma_{\gamma}(x,y,t) &= (\nu -1)\Sigma_f + \begin{cases}
																		s(t), & (x,y)\in \Omega_R, \\
																		0, &\tx{else},
																		\end{cases}
}
where the domain $\Omega_R$ is defined as
\balign{
\Omega_R = \{ (x,y)\in\R^2: -0.5 &\leq x\leq 0.2, \, -0.5\leq y \leq 0, \\
											  -0.3 &\leq x\leq 0, \, 0<y \leq 0.6, \\
												  0 &<x\leq 0.5, \, 0<y<0.3  \}.
}
The reason for the complicated design of $\Omega_R$ is that a simple rectangular shape would not produce spatial gradients which are large enough to distinguish between $P_3$ and $SP_3$ solutions.
\begin{figure}
 \centering
 \subfloat[Rod Geometry]{\includegraphics[scale=0.4]{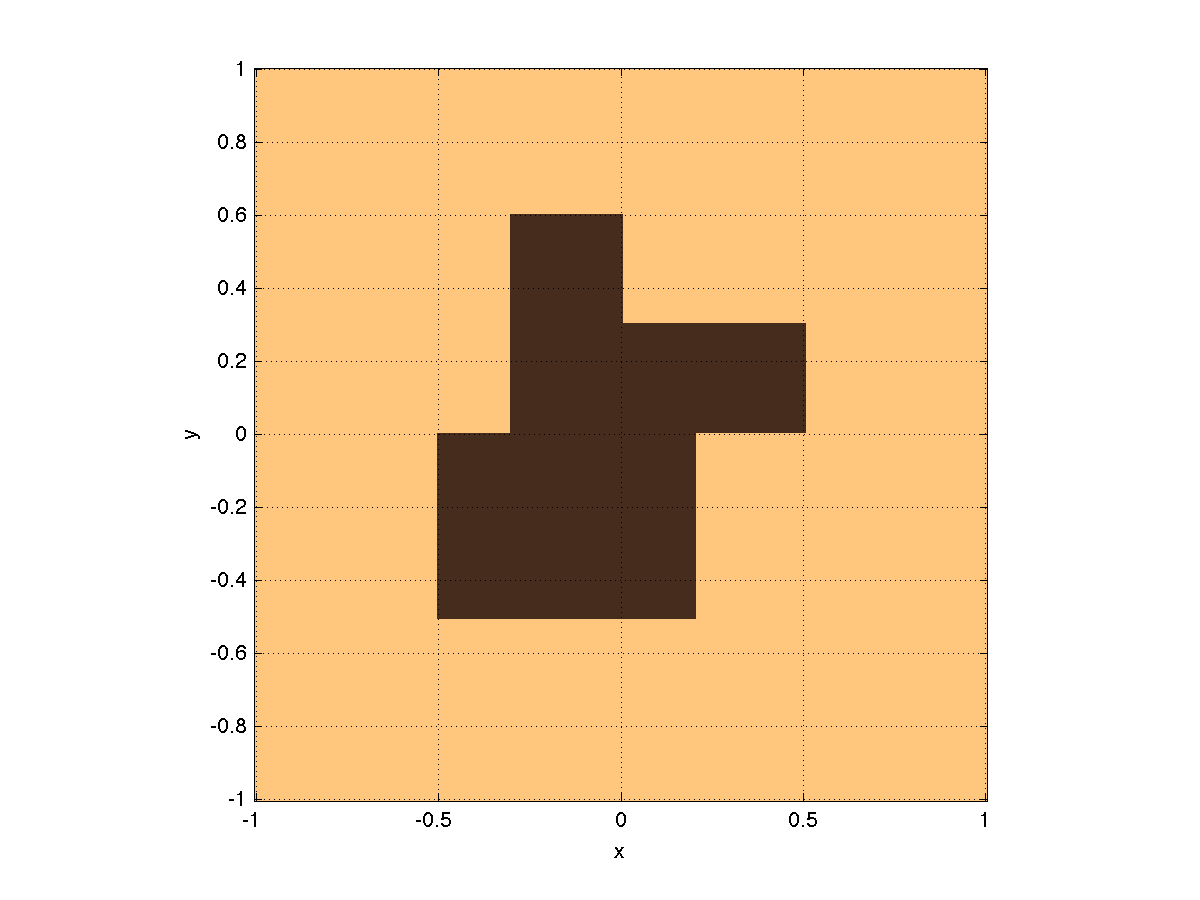}}
 \,
 \subfloat[$P_{39}$ solution at $t=0.3$: $s_{\tx{max}} = 10, \Delta T=0.2$]{\includegraphics[scale=0.6]{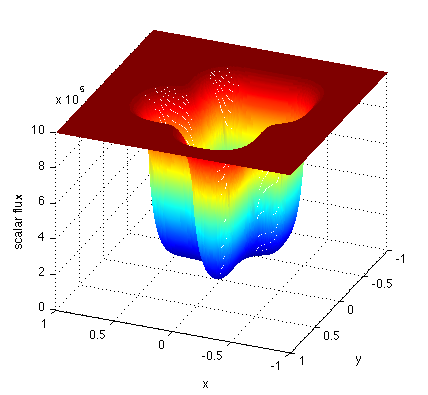}}
 \caption{}
  \label{pict:ControlRod}
 \end{figure}

This definition of $\Sigma_{\gamma}$ models an asymmetric rod with a cross section shown in \reffig{pict:ControlRod}a, which is moved into or out of the moderator material in a way specified by the function $s(t)$.
We choose a sequence of three processes:
\begin{itemize}
\item{pushing the rod into the material over
the time of $\Delta T$,}
\item keeping it in the moderator for the time of $T_{\tx{max}}-2\Delta T$ and
\item pulling the rod out over
the same time of $\Delta T$,
\end{itemize}
and, hence, define
\balign{
s(t) = \begin{cases}
		 \frac{s_{\tx{max}}}{\Delta T} \, t, & 0< t \leq \Delta T, \\
		 s_{\tx{max}}, & \Delta T < t < T_{\tx{max}}-\Delta T, \\
		 \frac{s_{\tx{max}}}{\Delta T} \, (T_{\tx{max}} -t), & T_{\tx{max}} -\Delta T \leq t \leq T_{\tx{max}}.
		\end{cases}
}

$T_{\tx{max}}$ is the duration of the whole process and is set to $T_{\tx{max}}=0.6$.
The quantity $s_{\tx{max}}$ determines how deep the rod is pushed into the moderator and can be associated with the maximum penetration depth. 
The ratio $s_{\tx{max}}/\Delta T$ models the speed of motion of the rod.
For large values of $s_{\tx{max}}/\Delta T$ the rod is moved very quickly, and thus large time derivatives are generated.
Due to the finite geometry of the rod, large gradients in space additionally occur.

The main goal of this problem is to analyze the behavior of diffusion, $SP_N$ and $P_N$ solutions in 2D for large time and space gradients in a semi-realistic setting where the absorption cross section depends on all three variables $x,y$ and $t$.

\begin{figure}
 \centering
 \subfloat[$t=0.3$]{\includegraphics[scale=0.9]{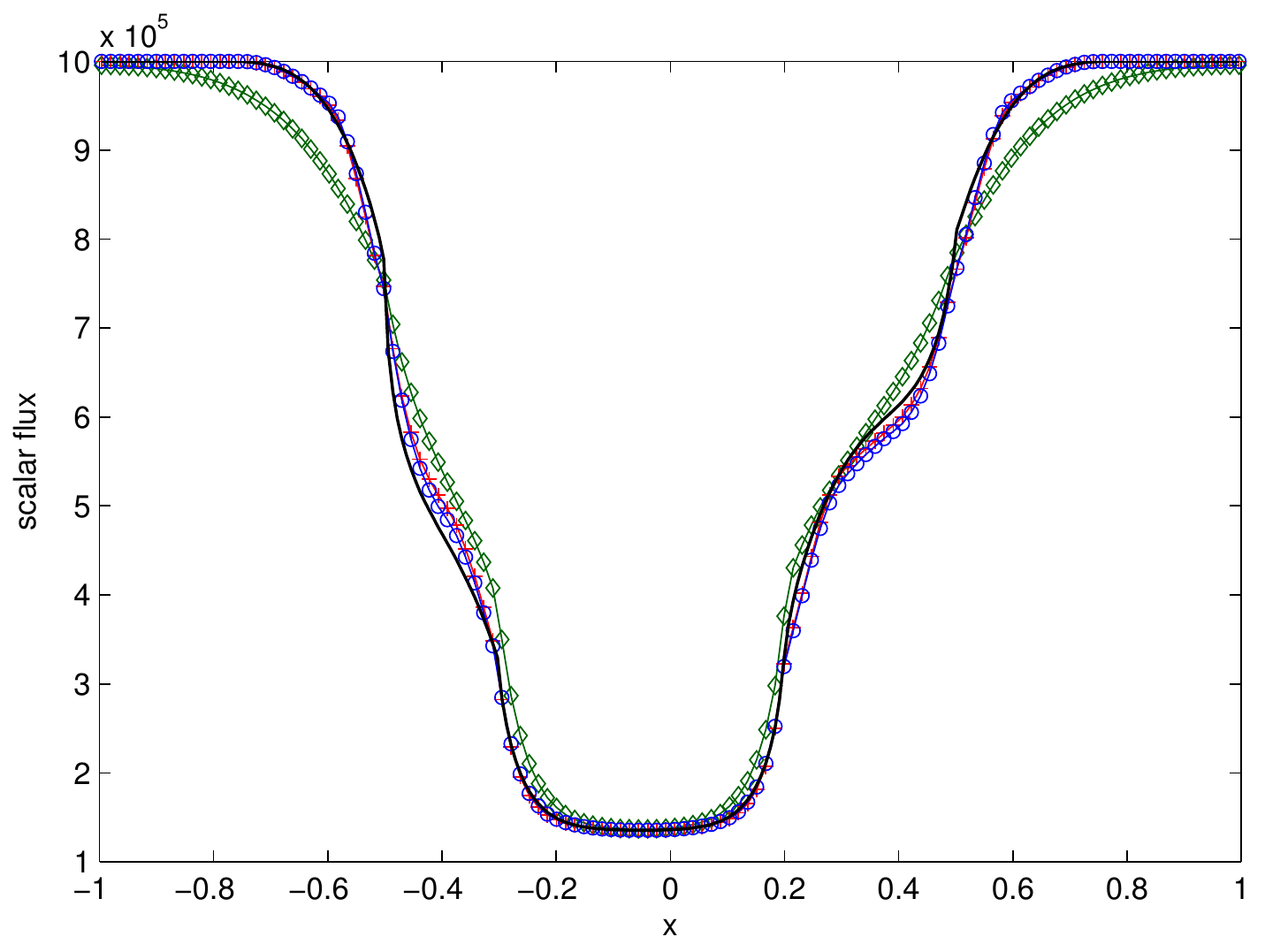}}
 \,
 \subfloat[$t=1$]{\includegraphics[scale=0.9]{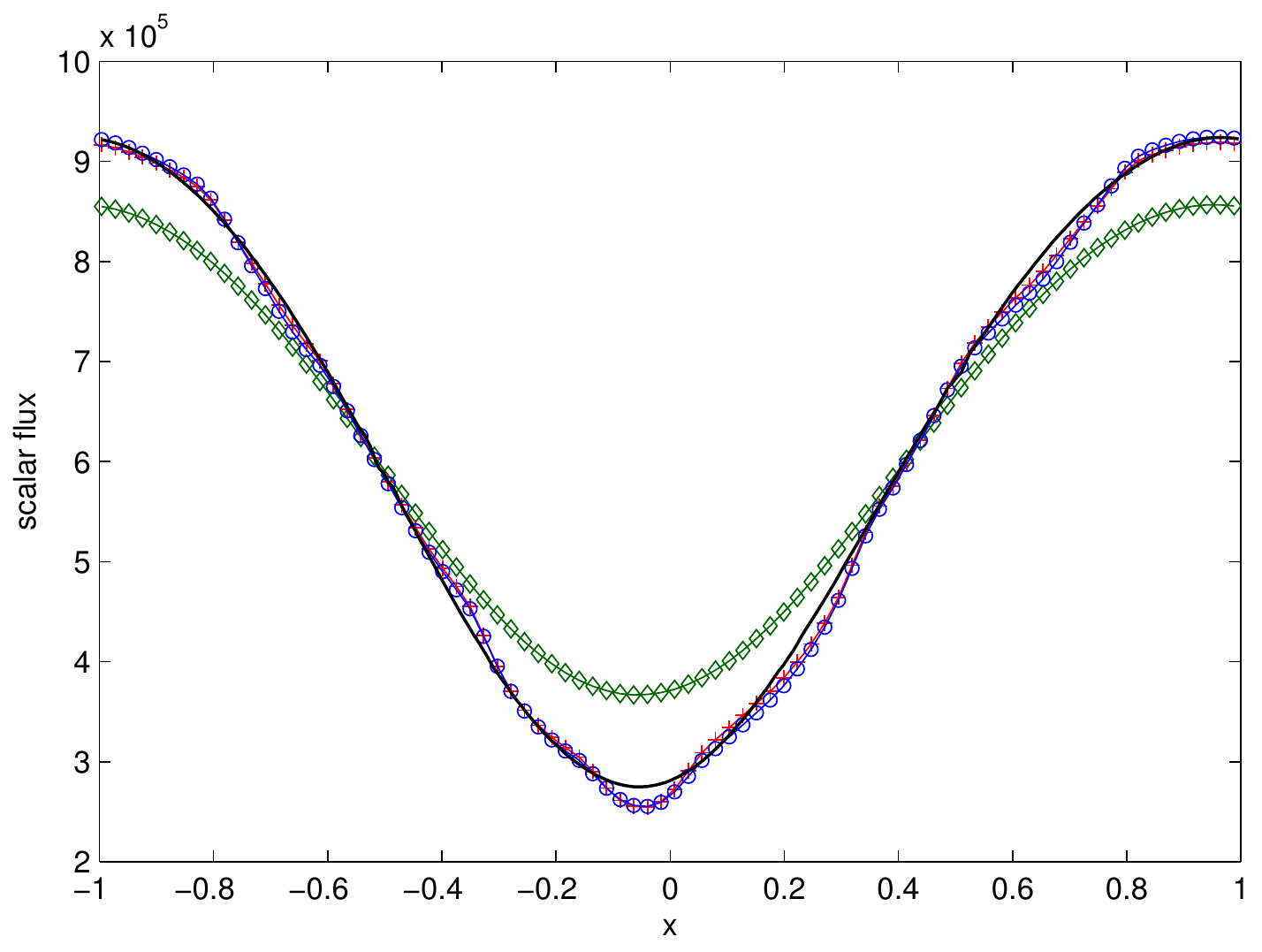}}
 \,
\caption{$\phi_0$ along $y=0$, $s_{\tx{max}} = 10, \Delta T=0.2$: Diffusion (green diamond line), $SP_3$ (red cross line), $P_3$ (blue circle line), $P_{39}$ (black solid line).}
  \label{pict:ControlRodGentle}
 \end{figure}

Figure~\ref{pict:ControlRodGentle} displays a cut of the scalar flux distribution along $y=0$ for $251\times 251$ discretization points.
The $P_{39}$ solution is considered as our reference, which can be seen in \reffig{pict:ControlRod}.
When the control rod is moved into the moderator, neutrons are absorbed and hence, their number diminishes.
If this is done at a small speed (compared to the velocity of the particles), neutrons will quickly flood the region close to the absorber.
Therefore, the scalar flux (which is a quantity for the particle number) is smooth but still forms steep slopes in our setting (\reffig{pict:ControlRodGentle}).
Although the solution drastically decreases, both the $SP_3$ and $P_3$ approximations are close to the reference $P_{39}$ result.
Moreover, $P_3$ is very close to $SP_3$.
On the other hand, the diffusion solution is inaccurate in regions where large spatial gradients are formed.
For increasing time, after the rod has been pulled out of the system, the neutrons spread in the whole domain and their distribution flattens more and more until, in steady-state, $\phi_0$ levels off to a constant value which is smaller than at the beginning of the process.
\reffig{pict:ControlRodGentle}b shows the distribution at an advanced time, where $P_3$ and $SP_3$ are still accurate and the diffusion approximation is far off the reference solution.


Our observations described above are consistent with the analysis in Section \ref{sec:asympDer}.
The asymptotic analysis implies that the $SP_3$ approximation is a higher-order correction to diffusion in cases where steep spatial slopes occur.
As long as these slopes are not extremely large, $SP_3$ and $P_3$ approximations yield similar results.
However, when a certain threshold is reached, the problem significantly leaves the asymptotic limit.
As a consequence, differences between $SP_3$ and $P_3$ become larger and $SP_3$ might lose accuracy.
%
%

\subsection{Checkerboard}
\label{subsec:checkerboard}

\begin{figure}
\centering
\includegraphics[scale=0.7]{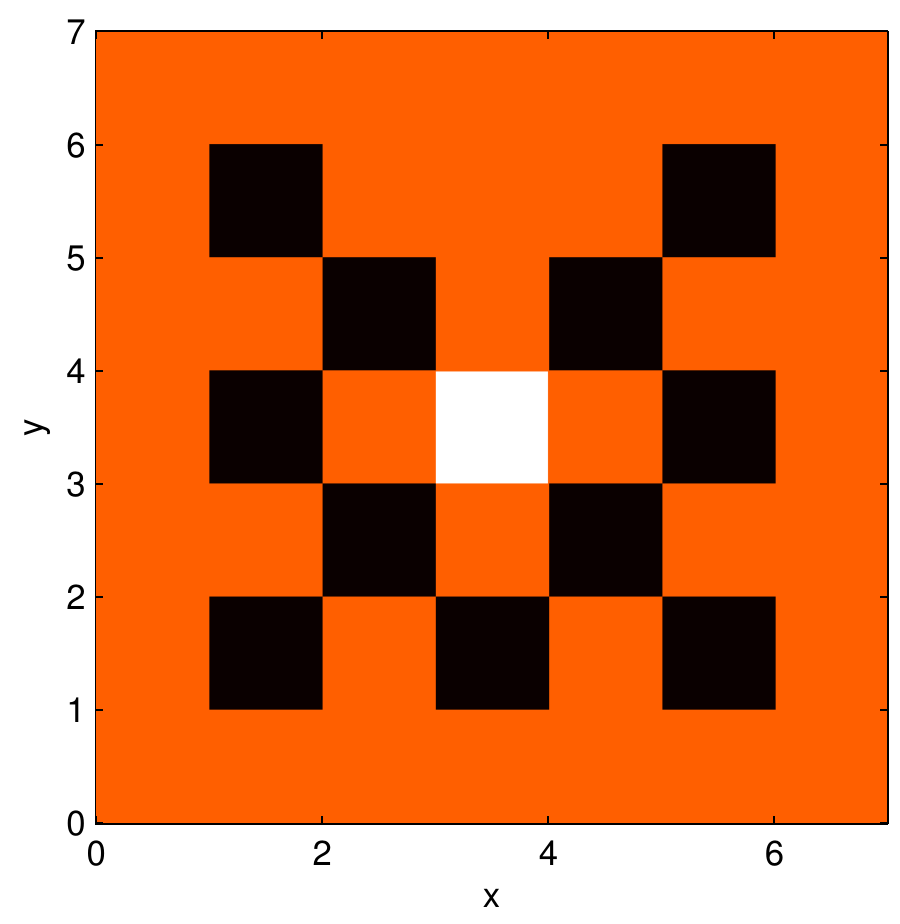}
\caption{Checkerboard test problem. Material coefficients: isotropic source (white), purely scattering $\sigs=1=\sigt$ (orange and white), purely absorbing $\siga=10=\sigt$ (black).}
\label{pict:checkbrd1}
\end{figure}

We consider the checkerboard problem from \cite{Bru02}:
The domain is a square of size $[0,7]\times [0,7]$ where the majority of the region is purely scattering.
In the middle of the lattice system, in the square $[3,4]\times [3,4]$, an isotropic source $Q=1$ is continuously generating particles.
Additionally, there are eleven small squares of side length one of purely absorbing spots in which $\siga=10=\sigt$.
\reffig{pict:checkbrd1} illustrates the problem settings more precisely.
Vacuum boundary conditions \cite{Bru02} are enforced and all initial quantities at $t=0$ are zero.
We compare the scalar flux using different methods (including high-order $P_N$ and $SP_N$).

\begin{figure}
\centering
\subfloat[Diffusion]{\includegraphics[scale=0.75]{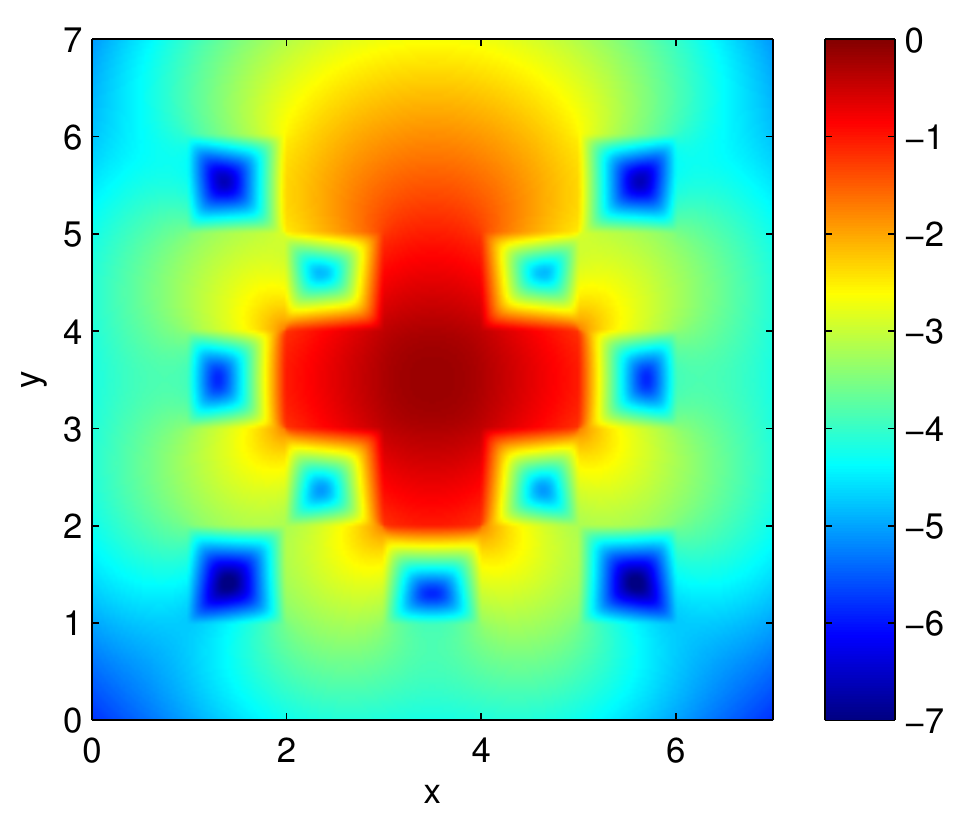}}
\,
\subfloat[$P_{39}$]{\includegraphics[scale=0.75]{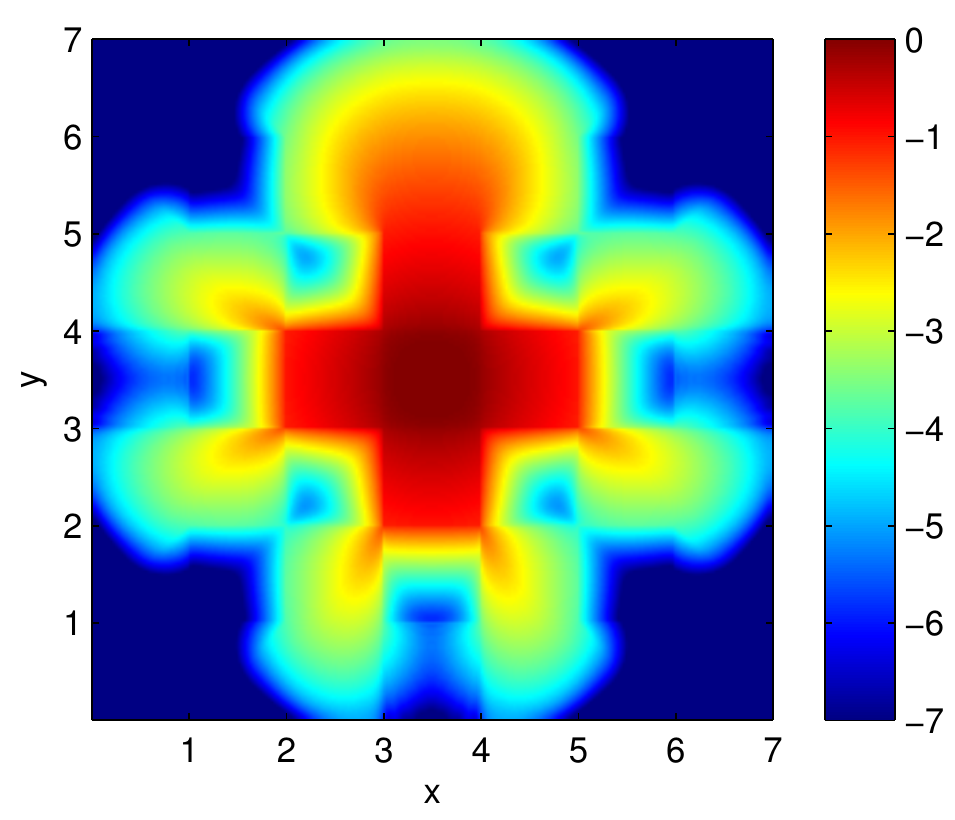}}
\,
\subfloat[$SP_3$]{\includegraphics[scale=0.75]{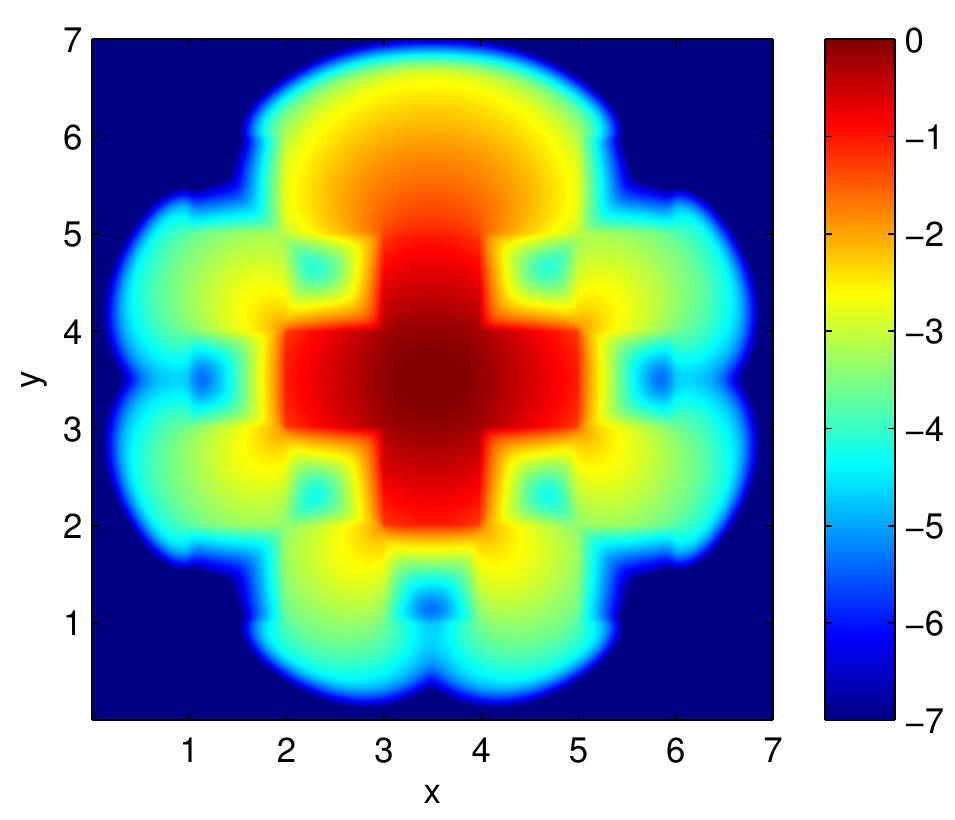}}
\,
\subfloat[$P_3$]{\includegraphics[scale=0.75]{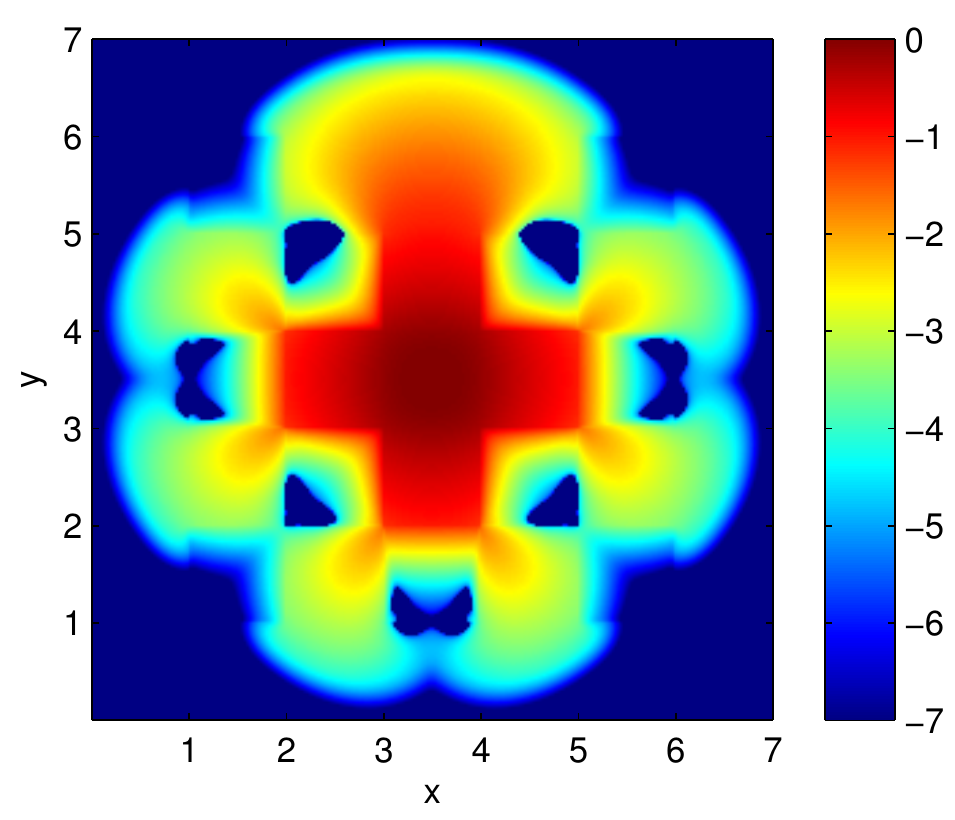}}
\,
\subfloat[$SP_{5}$]{\includegraphics[scale=0.75]{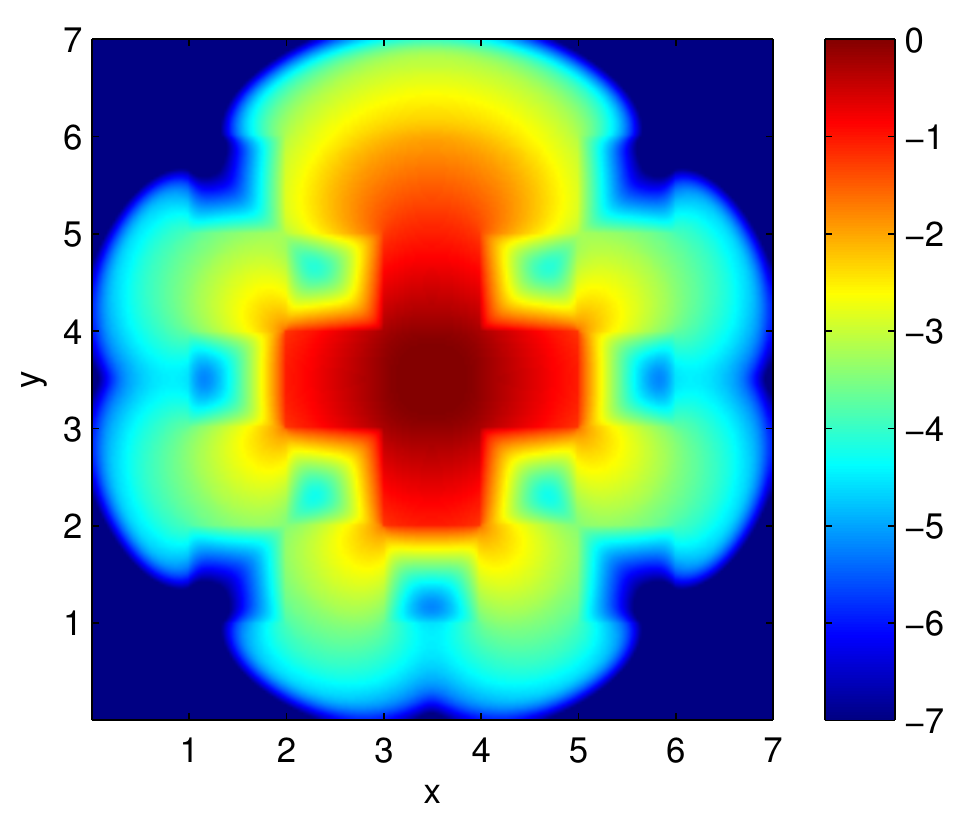}}
\,
\subfloat[$P_{5}$]{\includegraphics[scale=0.75]{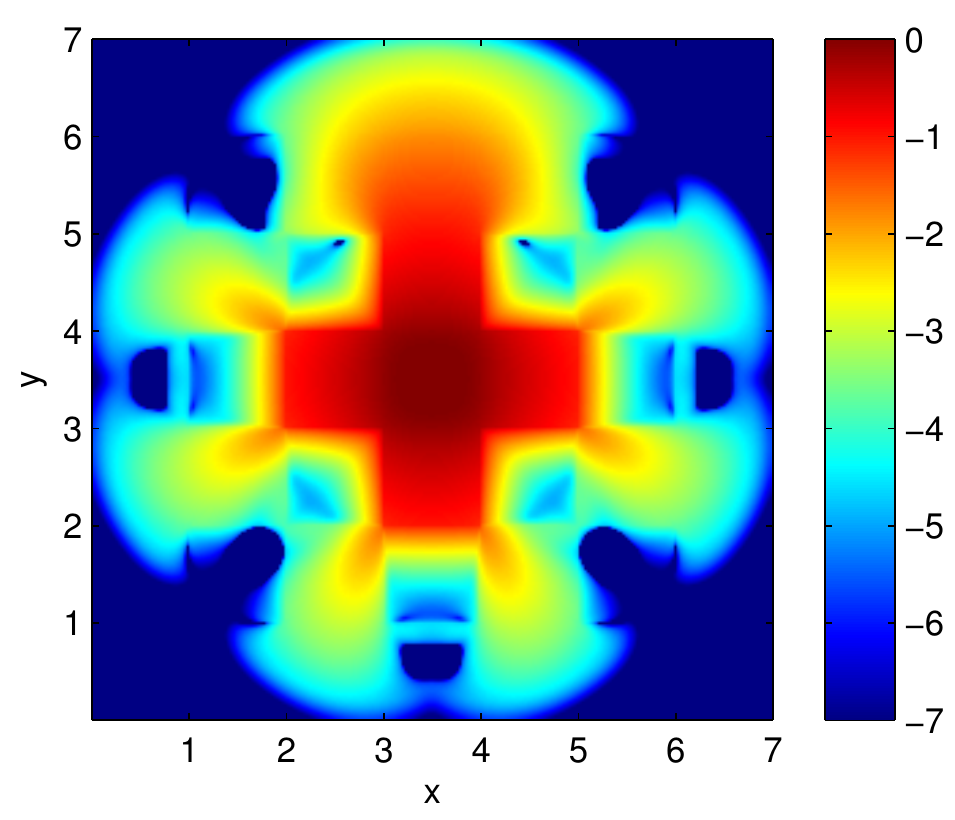}}
\caption{Checkerboard problem: Scalar flux $\phi_0$ at $t=3.2$ for from several approximations with $250\times 250$ points. The values are plotted in $\log_{10}(\phi_0)$ and limited to seven orders of magnitude.}
\label{pict:checkbrd2}
\end{figure}

The diffusion solution in \reffig{pict:checkbrd2} gives a poor result.
Particles are transported from the central source at a much higher speed to the boundaries of the domain.
The solution is much smoother than it should be, and large slopes are washed out.

Although the $SP_3$ calculation shows large improvements upon the diffusion result, it is still too diffusive compared to the reference $P_{39}$ (\reffig{pict:checkbrd2}).
In some purely absorbing regions away from the center, the $P_3$ computation lacks particles because the approximation does not allow particle waves to travel at a high speed.
Hence, depending on the desired purpose, $P_3$ is not evidently superior to $SP_3$ in this case.
Nevertheless, the particle beams between the corners of the absorbing regions are well resolved in all $P_N$ solutions.

\begin{figure}
\centering
\includegraphics[scale=0.95]{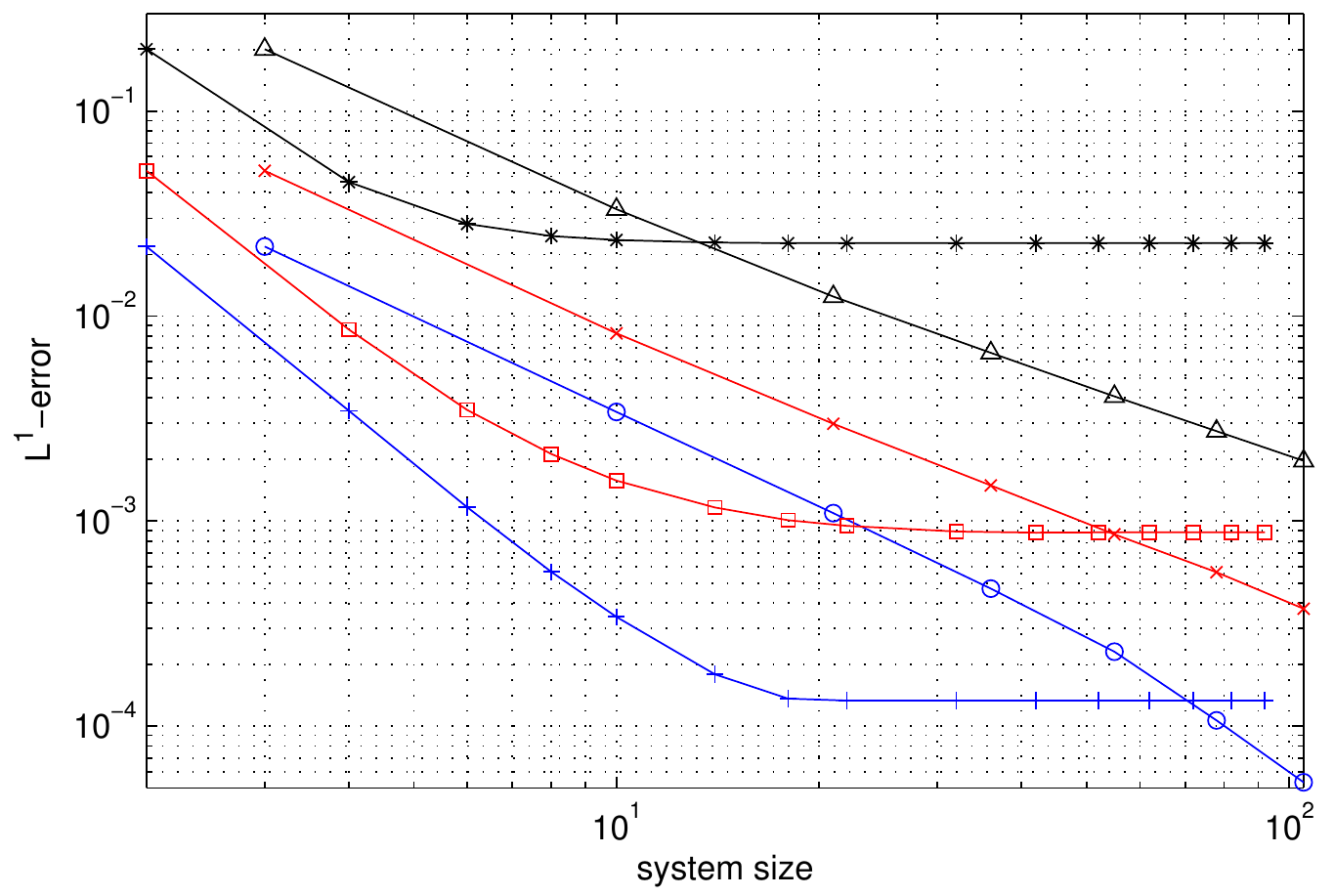}
\caption{$L^1$-error between $SP_N$ and $P_{39}$ or $P_N$ and $P_{39}$ for $1000\times 1000$ spatial discretization points: $\eps=0.5: P_N$ (black solid triangle line), $\eps=0.5: SP_N$ (black solid asterisk line), $\eps=0.1: P_N$ (red solid cross line), $\eps=0.1: SP_N$ (red solid square line), $\eps=0.04: P_N$ (blue solid circle line), $\eps=0.04$ (blue solid plus line).}
\label{pict:checkbrd3}
\end{figure}

For increasing $N$, the spherical harmonics solutions show better improvements than $SP_N$.
Although $SP_5$ calculation yields visible changes, it still shows obvious differences from the reference.
It is important to emphasize that $SP_N$ solutions of this problem cannot be expected to be highly accurate for several reasons:
The underlying cross-sections are discontinuous and lead to a transport solution with steep slopes varying over at least seven orders of magnitude.
Hence, the scaling parameter $\eps$ from Section \ref{sec:asympDer} cannot be small.
Moreover, purely absorbing regions have a scattering ratio of $c=\frac{\sigs}{\sigt}=0$ which is significantly outside any asymptotic limit discussed in Section \ref{sec:asympDer}.
And last, the total cross-section $\sigt$ depends on both spatial variables $x$ and $y$, which violates an assumption in our asymptotic derivation.

However, realistic applications often include these kinds of geometries, in which cross-sections are (highly) varying and entail large spatial gradients in solutions.
If $SP_N$ approximations are computed for these problems, it will be of significant interest to know how much is gained by increasing order $N$.
In other words, for growing $N$, how much does the error between $SP_N$ and transport solution decrease?
An answer to this question is necessary to determine the benefit of a high-order $SP_N$ calculation in comparison to the additional computational effort.

To demonstrate the behavior of this error in a numerical way, $SP_N$ approximations are calculated for odd $N$, from $N=1$ to $N=91$.
In order to make the numerical discretization error comparabely small we choose a very fine discretization grid of $1000\times 1000$ points. 
Additionally, a scaling parameter $\eps$ is introduced into all equations to achieve the scaling from \refeq{asympDer:eq2c}.
Approximative solutions are then computed for the scaled transport equation (\ref{asympDer:eq8}) where the velocity $v$ is set to one.
Again, $P_{39}$ is assumed to be the reference solution.

\reffig{pict:checkbrd3} shows the $L^1$-error between $SP_N$ and $P_{39}$ or $P_N$ and $P_{39}$ for different values of $\eps=0.04, 0.1, 0.5$.
The error is plotted against the system size (i.e., the number of equations to be solved) which is $N+1$ for the $SP_N$ and $(N+1)(N+2)/2$ for the $P_N$ approximation.
This choice is made to give a reasonable comparison based on computational effort between $P_N$ and simplified $P_N$ calculations. 

Indeed, both the $P_N$- and $SP_N$-error become smaller for decreasing values of $\eps$.
Moreover, for a fixed $\eps$ and increasing order $N$, the error of the $P_N$ and $SP_N$ method decreases monotonically.
The $P_N$-error is even strictly decreasing to zero.
In contrast, due to the discontinuous property of the cross sections or a scattering ratio which is outside the asymptotic limit, the $SP_N$-error approaches a constant non-zero value and saturates at system sizes of roughly $10-20$.

For a fixed system size (i.e. the same number of equations to be solved), \reffig{pict:checkbrd3} shows $SP_N$-errors which are smaller than the corresponding $P_N$-errors.
Consequently, to achieve a desired $L^1$-error above a minimum threshold the $SP_N$ approximation is computationally less costly than $P_N$ calculations.
However, this is only valid for small system sizes and errors which are above the saturation limit of the $SP_N$ method.

The asymptotic analysis in Section~\ref{sec:asympDer} is strictly valid for homogeneous media.
Nevertheless, the $SP_N$ error function for a small $\eps=0.04$ in \reffig{pict:checkbrd3} demonstrates that there is a significant error decrease for high order $SP_N$ solutions.
Hence, this behavior shows numerically that $SP_N$ computations of large orders $N\geq 5$ can also be used in settings with discontinuous cross sections to achieve a specified error.
\section{Discussion}
\label{sec:disc}

We have presented an approach on how time-dependent $SP_N$ equations can be derived in an ad hoc way for general $N$, and have explicitely provided an asymptotic analysis for these equations up to $N=3$.
A crucial result is the system of $SP_3$ equations from \refeqs{clDer:eq15}, a first-order system of hyperbolic PDEs.
One feature of \refeqs{clDer:eq15} is that in 1-D planar geometry, they reduce \textit{exactly} to the time-dependent 1-D planar-geometry $P_3$ equations.
Hence, \textit{in planar geometry}, numerical solutions of the time-dependent $SP_3$ and $P_3$ equations are equivalent.
Numerical calculations are therefore performed in two space dimensions in Section \ref{sec:numres}.
Moreover, computations in Section \ref{sec:PnSpnEquiv} demonstrate that there are even cases outside the asymptotic limit where $P_N$ and $SP_N$ are equivalent in 2D.

Problems in heterogeneous media with cross sections of small scattering ratios $c=\frac{\sigs}{\sigt}$ show the following behavior:
The smaller the scaling parameter $\eps$ becomes the closer are $SP_N$ computations to transport solutions. 
On the other hand, large $\eps$ lead to large errors (Section \ref{subsec:checkerboard}).
Solutions to problems which do not satisfy the homogeneity assumption, made in the asymptotic analysis, still show accurate solutions (Section \ref{subsec:checkerboard}).
Furthermore, all comparisons confirm that $SP_3$ results are throughout superior to diffusion solutions, especially in cases where large gradients are formed. \\ \\
In the following, a few possible future tasks are discussed:

\begin{enumerate}[1.]
\item{
The asymptotic analysis in this paper gives rise to time-dependent $SP_3$ equations which differ from those developed in \cite{FraKlaLar07}.
From a theoretical point of view, the main reason is the different scaling, which especially yields an additional $\eps^2$ in front of the time-derivative in \cite{FraKlaLar07}.
Amongst others, one consequence of the different scaling is that the previous time-dependent $SP_3$ equations do not reduce \textit{exactly} to the time-dependent planar-geometry $P_3$ equations.
Another difference is that the time-dependent $SP_3$ theory in \cite{FraKlaLar07} contains an arbitrary constant $\alpha$.
Although all admissible values of $\alpha$ imply the same asymptotic order of accuracy, it could be still valuable to compare numerical solutions of those equations with numerical solutions of equations proposed in this publication.
Central challenges are to explain different behaviors of these equations for certain problems and to verify whether computational results confirm the theoretical predictions.
}
\item{
Recently, Larsen presented \textit{modified} diffusion and $SP_N$ equations \cite{Lar11a}, which are designed to be more accurate for deep penetration problems.
The asymptotic analysis therein covers the steady-state, anisotropically scattering linear Boltzmann equation and is based on a different scaling.
Can a similar procedure be applied to the time-dependent $SP_N$ equations considered in this paper?
Answers to this question could provide a mathematical foundation for the time-dependent $SP_N$ equations in a much more general field of applications.
}
\item{
It is difficult to calculate exact angular fluxes of the transport equation \cite{Gan08} and obviously, even more difficult to compare them to approximations.
Nevertheless, it is still possible to gain a different type of information about the analytic solution.
One possibility to investigate the accuracy of angular moment approximations to the Boltzmann equation is to perform a moment analysis.
The main idea of this method is to calculate angular and/or spatial moments of the angular flux and compare them to corresponding quantities in the moment equations.
Consequently, one can draw conclusions about the accuracy of the method. 
Densmore and McClarren \cite{DenMcC11}, e.g., compared the previous time-dependent simplified $P_N$ methods from \cite{FraKlaLar07} to other approximations.
A similar analysis would provide additional information about moments which are preserved by the time-dependent $SP_N$ solutions from this paper.
}
\item{
McClarren suggests one important aspect about the $SP_N$ equations to be studied in the future \cite{McC11}:
\emph{What is the optimal order of $SP_N$ equations that determines a limit where your benefit in accuracy is still larger than the additional computational costs?} Or in other words, up to which order $N$ is it reasonable to use the $SP_N$ equations? No investigations concerning this issue have been performed up to now.
However, the publicly available MATLAB code \cite{SeiboldFrank2012} used in this paper allows to perform $P_N$ and $SP_N$ calculations in 2D of very high order.
In Section \ref{subsec:checkerboard} the error behavior between $SP_N$ and the transport solution is studied for increasing order and different scaling parameters. The results indicate that up to a comparatively high order around $N=9$ the $SP_N$ equations are superior to the $P_N$ equations of same system size. These results could be a motivation for future analysis.
}
\item{
As we only considered mono-energetic transport problems here, the extension to multi-group calculations along the lines of \cite{LarMorMcG96} will allow for the simulation of realistic applications and is left to future work.
}
\end{enumerate}

\begin{appendix}
\section{Ad Hoc Derivation: Time-Dependent $SP_N$ equations of Arbitrary Order}
Although a mathematical foundation is only given for the $SP_3$ equations in this paper, 
we also perform numerical computations with equations of higher order.
Therefore, we briefly explain how time-dependent $SP_N$ equations of arbitrary order can be obtained.

Starting from the 1-D time-dependent $P_N$ equations in planar geometry, the following hyperbolic system of $SP_N$ equations can be derived by a similar procedure from Section~\ref{sec:clDer}:
\seqalign{\label{appA:eq1}}{
\recv \D{\phi_{2i}}{t} (\ux,t) &+ \unab\cdot \ul{J}_{2i} (\ux ,t)+\sigt(\ux,t)\phi_{2i}(\ux,t) = \delta_{i,0}( \sigs \phi_0(\ux,t) +Q(\ux,t) ), \\
\recv\D{\ul{J}_{2i}}{t} (\ux ,t) &+ \unab  \left \{k_i \phi_{2i+2}(\ux,t) + l_i \phi_{2i}(\ux,t) + m_i \phi_{2i-2}(\ux,t) \right \} +\sigt(x,t) \ul{J}_{2i}(\ux, t),
}
where $i=0,1,\ldots, \lfloor \frac{N}{2} \rfloor$ and $\lfloor \, \, \rfloor$ is the floor function.
The coefficients in \refeqs{appA:eq1} are given by
\seqalign{}{
k_i &= \frac{2i(2i-1)}{(16i^2-1)}, \\
l_i &= \begin{cases}
\ds \frac{32i^3+24i^2-1}{(16i^2-1)(4i+3)}, \quad & N \tx{ odd}, \\
\ds \frac{4i^2}{(16i^2-1)}, \q & N \tx{ even},
\end{cases} \\
m_i &= \begin{cases}
\ds \frac{2(2i^2+3i+1)}{(4i+1)(4i+3)}, \q & 2(i+1) \leq N, \\
0, & \tx{else}.
\end{cases}
}
%
%
%
%
\section{Expanding Operators}

Here, we provide detailed algebraic manipulations for the simplification of the expansion operators

\balign{ 
{\mathcal L}_n = \frac{1}{4\pi} \int_{S^2} \uomega\cdot\unab ( \momegax )^n d\Omega,
} 
where
\balign{
{\mathcal T} = \frac{1}{v\Sigma_t} \D{}{t}, \q \ul{{\mathcal X}} = \frac{1}{\sigt} \unab, \q {\mathcal P} = \frac{1}{4\pi} \int_{S^2} (\cdot) \, d\Omega \quand {\mathcal M} = \invOp (I- {\mathcal P}).
}

If we denote the cosine of the polar angle by $-1 \leq\mu\leq 1$ and the azimuthal angle by $\q 0\leq\phi < 2\pi$, then the unit vector $\ul{\Omega} = (\Omega_1,\Omega_2,\Omega_3)$ on the unit sphere is given by its scalar components
\balign{
\Omega_1 &= \mu,  \\
\Omega_2 &= \sqrt{1-\mu^2} \cos(\phi), \\
\Omega_3 &= \sqrt{1-\mu^2} \sin(\phi).
}
Much of the following analysis is based on results of the integral
\[ \int_{S^2} \Omega_{i_1}\Omega_{i_2}\ldots\Omega_{i_n} d\Omega  \] 
for all possible combinations of $1\leq i_1,\ldots,i_n \leq 3$. 
Without a proof, we explicitly state some solutions which are needed later on:
\seqalign{\label{app:eq1}}{
\int_{S^2} \Omega_{i_1}\Omega_{i_2} \, d\Omega &= \frac{4\pi}{3} \delta_{i_1,i_2}, \label{app:eq1a}\\
\int_{S^2} \Omega_{i_1}\Omega_{i_2}\Omega_{i_3}\Omega_{i_4} \, d\Omega &= \frac{4\pi}{15} \left ( \delta_{i_1,i_2}\delta_{i_3,i_4} +\delta_{i_1,i_3}\delta_{i_2,i_4} +\delta_{i_1,i_4}\delta_{i_2,i_3} \right ), \label{app:eq1b} \\
\int_{S^2} \Omega_{i_1}\Omega_{i_2}\ldots\Omega_{i_n} \, d\Omega &= 0 \q\text{for odd } n.  \label{app:eq1c}
}
We first focus on calculating ${\mathcal L}_n$ for even $n$ because above integral is zero for odd $n$:
\balign{
{\mathcal L}_0 &= \frac{1}{4\pi} \int_{S^2} \uomega\cdot\unab \, d\Omega = 0.
}
\balign{
{\mathcal L}_2 &= \frac{1}{4\pi} \int_{S^2} \uomega\cdot\unab  ( \momegax )^2 \, d\Omega \\
&= \frac{1}{4\pi} \int_{S^2} \uomega\cdot\unab (\momegax)(\momegax)  \, d\Omega \\
&=  \frac{1}{4\pi} \int_{S^2} \uomega\cdot\unab (\momegax) \invOp (I-{\mathcal P}) \omegax d\Omega.
}
Since
\balign{
{\mathcal P} (\uomega\cdot {\mathcal \ul{X}} ) \phix = \frac{1}{\sigt} \sum_{i=1}^3 \partial_i \phix \int_{S^2} \Omega_i \, d\Omega = 0, \label{app:eq2}
}
it follows
\balign{
{\mathcal L}_2 &= \frac{1}{4\pi} \int_{S^2} \uomega\cdot\unab (\momegax) \invOp d\Omega \\
&= \frac{1}{4\pi} \int_{S^2} \uomega\cdot\unab\invOp (I-{\mathcal P}) \omegax \invOp\omegax d\Omega \\
&= \frac{1}{4\pi} \int_{S^2} \uomega\cdot\unab\invOp \{ \omegax \invOp\omegax \nonumber  \\
& \hspace{6cm} -{\mathcal P} \omegax \invOp\omegax \} \, d\Omega \label{app:eq3} \\
&= \frac{1}{4\pi} \int_{S^2} \uomega\cdot\unab\invOp \omegax \invOp\omegax d\Omega \\
&= 0,
}
where last two equalities follow from the fact that
\balign{
\left [ {\mathcal P} \omegax \invOp\omegax \right ] \phix
} 
is independent of $\uomega$. Hence, only integrals of an odd number of $\uomega$ are left which all vanish according to \refeq{app:eq1c}.
Similarly, we obtain
\balign{{\mathcal L}_4 &= 0, \\
{\mathcal L}_6 &= 0,
}
and attend to odd-numbered operators ${\mathcal L}_n$: Note that \refeq{app:eq2} yields
\balign{
\uomega\cdot\unab (\momegax) &= \uomega\cdot\unab \invOp (I -{\mathcal P} ) \frac{1}{\sigt} \uomega\cdot\unab \nonumber \\
&= \uomega\cdot\unab \invOp \frac{1}{\sigt} \uomega\cdot\unab.
}
Using \refeq{app:eq1a}, we obtain
\balign{
{\mathcal L}_1 =  \frac{1}{4\pi}  \int_{S^2} \uomega\cdot\unab (\momegax) d\Omega &=  \frac{1}{4\pi}  \sum_{i_1,i_2=1}^3 \int_{S^2} \Omega_{i_1} \partial_{i_1} \invOp\frac{1}{\sigt}\Omega_{i_2}\partial_{i_2} \, d\Omega \label{app:eq11a} \\
&=  \frac{1}{4\pi}  \sum_{i_1,i_2=1}^3 \frac{4\pi}{3} \delta_{i_1,i_2} \left ( \partial_{i_1} \invOp\frac{1}{\sigt}\partial_{i_2} \right )  \label{app:eq11b} \\
&=   \frac{1}{3} \sum_{i_1=1}^3 \partial_{i_1} \invOp\frac{1}{\sigt}\partial_{i_1} \\
&= \frac{1}{3} \unab\cdot \invOp \ul{{\mathcal X}} \label{app:eq11d}.
}
Next, making use of \refeq{app:eq1c} we calculate
\balign{
{\mathcal L}_3 &=  \frac{1}{4\pi}  \int_{S^2} \uomega\cdot\unab (\momegax)(\momegax)(\momegax) d\Omega \\
&= \frac{1}{4\pi}  \int_{S^2} \uomega\cdot\unab (\momegax) (\momegax) [\invOp (I -\calP)\uomega\cdot\ul{\calX}  ]  d\Omega \\
&= \frac{1}{4\pi}  \int_{S^2} \uomega\cdot\unab (\momegax) [\invOp (I -\calP)\uomega\cdot\ul{\calX}  ] [\invOp \uomega\cdot\ul{\calX}  ]  d\Omega \\
&= \frac{1}{4\pi}  \int_{S^2} \uomega\cdot\unab (\momegax) [\invOp \uomega\cdot\ul{\calX}  ] [\invOp \uomega\cdot\ul{\calX}  ]  d\Omega \nonumber \\
& \q -\frac{1}{4\pi}  \int_{S^2} \uomega\cdot\unab (\momegax) \invOp\calP \left \{ \uomega\cdot\ul{\calX} \invOp \uomega\cdot\ul{\calX} \right \}  d\Omega. \label{app:eq12}
}
According to \refeq{app:eq1c}, last integral simplifies to
\balign{
&-\frac{1}{4\pi}  \int_{S^2} \uomega\cdot\unab (\momegax) \invOp\calP \left \{ \uomega\cdot\ul{\calX}\invOp \uomega\cdot\ul{\calX} \right \}  d\Omega \\
=&- \frac{1}{4\pi}  \int_{S^2} \uomega\cdot\unab [\invOp \uomega\cdot\ul{\calX}  ] \invOp\calP \left \{ \uomega\cdot\ul{\calX} \invOp \uomega\cdot\ul{\calX} \right \}  d\Omega \nonumber \\
&+ \frac{1}{4\pi}  \int_{S^2} \uomega\cdot\unab \invOp \underbrace{ \calP \left \{ \uomega\cdot\ul{\calX} \invOp \calP  \left \{ \uomega\cdot\ul{\calX}  \invOp \uomega\cdot\ul{\calX} \right \} \right \} }_{=0} d\Omega \\
=&- \frac{1}{4\pi}  \int_{S^2} \uomega\cdot\unab [\invOp \uomega\cdot\ul{\calX}  ] \invOp\calP \left \{ \uomega\cdot\ul{\calX} \invOp \uomega\cdot\ul{\calX} \right \}  d\Omega.
}
Similarly, we can eliminate the operator $\calP$ in the first term of \refeq{app:eq12} and obtain
\balign{
{\mathcal L}_3 &= \frac{1}{4\pi}  \int_{S^2} \uomega\cdot\unab [\invOp (I-\calP) \uomega\cdot\ul{\calX}  ] [\invOp \uomega\cdot\ul{\calX}  ] [\invOp \uomega\cdot\ul{\calX}  ]  d\Omega \nonumber \\
&- \frac{1}{4\pi}  \int_{S^2} \uomega\cdot\unab [\invOp \uomega\cdot\ul{\calX}  ] \invOp\calP \left \{ \uomega\cdot\ul{\calX} \invOp \uomega\cdot\ul{\calX} \right \}  d\Omega \\
&= \frac{1}{4\pi}  \int_{S^2} \uomega\cdot\unab [\invOp \uomega\cdot\ul{\calX}  ] [\invOp \uomega\cdot\ul{\calX}  ] [\invOp \uomega\cdot\ul{\calX}  ]  d\Omega \nonumber \\
&- \frac{1}{4\pi}  \int_{S^2} \uomega\cdot\unab [\invOp \uomega\cdot\ul{\calX}  ] \invOp\calP \left \{ \uomega\cdot\ul{\calX} \invOp \uomega\cdot\ul{\calX} \right \}  d\Omega \label{app:eq10}
}
We consider last two integral terms separately and expand the dot products therein:
\balign{
&\frac{1}{4\pi}  \int_{S^2} \uomega\cdot\unab [\invOp \uomega\cdot\ul{\calX}  ] [\invOp \uomega\cdot\ul{\calX}  ] [\invOp \uomega\cdot\ul{\calX}  ]  d\Omega \\
=& \frac{1}{4\pi}  \sum_{i_1,\ldots, i_4 = 1}^3 \int_{S^2} \Omega_{i_1} \partial_{i_1} \left ( \invOp \recsigmat \Omega_{i_2}\partial_{i_2} \right ) \nonumber \\
& \hspace{5cm} \cdot \left ( \invOp \recsigmat \Omega_{i_3}\partial_{i_3} \right ) \left ( \invOp \recsigmat \Omega_{i_4}\partial_{i_4} \right ) d\Omega \\
=&  \frac{1}{4\pi}  \sum_{i_1,\ldots, i_4 = 1}^3 \partial_{i_1} \left ( \invOp \recsigmat \partial_{i_2} \right )  \left ( \invOp \recsigmat\partial_{i_3} \right ) \left ( \invOp \recsigmat \partial_{i_4} \right ) \nonumber  \\
& \hspace{7cm} \cdot \int_{S^2} \Omega_{i_1}\Omega_{i_2}\Omega_{i_3}\Omega_{i_4} \, d\Omega \\
=& \sum_{i_1,\ldots, i_4 = 1}^3 \partial_{i_1} \left ( \invOp \recsigmat \partial_{i_2} \right )  \left ( \invOp \recsigmat\partial_{i_3} \right ) \left ( \invOp \recsigmat \partial_{i_4} \right ) \nonumber \\
& \hspace{7cm} \cdot  \frac{1}{15} \left ( \delta_{i_1,i_2}\delta_{i_3,i_4} +\delta_{i_1,i_3}\delta_{i_2,i_4} +\delta_{i_1,i_4}\delta_{i_2,i_3} \right ),
}
where last equality follows from \refeq{app:eq1b}.
Following the same arguments as in \refeqs{app:eq11a}-(\ref{app:eq11b}), the second integral term in \refeq{app:eq10} can be rewritten to
\balign{
&- \frac{1}{4\pi}  \int_{S^2} \uomega\cdot\unab [\invOp \uomega\cdot\ul{\calX}  ] \invOp\calP \left \{ \uomega\cdot\ul{\calX} \invOp \uomega\cdot\ul{\calX} \right \}  d\Omega \\
=& -  \sum_{i_1,\ldots, i_4 = 1}^3 \partial_{i_1} \left ( \invOp \recsigmat \partial_{i_2} \right )  \left ( \invOp \recsigmat\partial_{i_3} \right ) \left ( \invOp \recsigmat \partial_{i_4} \right ) \cdot \frac{1}{9} \delta_{i_1,i_2}\delta_{i_3,i_4}. \nonumber
}
Altogether, we get
\balign{
{\mathcal L}_3 =& \sum_{i_1,\ldots, i_4 = 1}^3 \partial_{i_1} \left ( \invOp \recsigmat \partial_{i_2} \right )  \left ( \invOp \recsigmat\partial_{i_3} \right ) \left ( \invOp \recsigmat \partial_{i_4} \right ) \nonumber \\
& \hspace{5cm} \cdot  \frac{1}{45} \left ( -2\delta_{i_1,i_2}\delta_{i_3,i_4} +3\delta_{i_1,i_3}\delta_{i_2,i_4} +3\delta_{i_1,i_4}\delta_{i_2,i_3} \right ). \label{app:eq13}
}
\begin{rem}
Equation~(\ref{app:eq13}) is exact and holds for arbitrary geometries.
In principle, \refeq{app:eq13} can be used to derive equations which are asymptotically valid in heterogeneous media.
However, due to the complexity of \refeq{app:eq13}, one would probably lose the simple structure of the $SP_N$ equations as presented in this paper.
To keep this main advantage of the $SP_N$ method further simplifications are needed and one possibility is shown in the following.
\end{rem}

\noindent Assuming that the medium is either
\begin{itemize}
\item one dimensional (where the sum of 81 terms reduces to one single term) or
\item homogeneous (where $\sigt$ is independent of the spatial variable $\ux$),
\end{itemize}
we can rearrange the derivative operators in \refeq{app:eq13} and conclude
\balign{
{\mathcal L}_3 =&\sum_{i_1,\ldots, i_4 = 1}^3 \frac{4\delta_{i_1,i_2}\delta_{i_3,i_4}}{45} \,\, \partial_{i_1} \left ( \invOp \recsigmat \partial_{i_2} \right )  \left ( \invOp \recsigmat\partial_{i_3} \right ) \left ( \invOp \recsigmat \partial_{i_4} \right ) \nonumber \\
=& \frac{4}{45} \left [ \unab\cdot  \invOp\ul{\calX} \right ] \invOp \left [ \ul{\calX} \cdot \invOp\ul{\calX} \right ].
}

A similar analysis can also be performed for the next operator in the hierarchy.
Since these expressions become even more involved and lengthy, we do not present them here.
A road map to this analysis is given in \cite{Lar11a} where a detailed derivation is also presented for ${\mathcal L}_3$.
Making the same assertions as above the operator simplifies to
\balign{
{\mathcal L}_5 &= \frac{44}{945} \left [\unab\cdot \invOp\ul{\calX} \right ] \invOp \left [ \ul{\calX}\cdot\invOp\ul{\calX}  \right ]  \invOp \left [ \ul{\calX}\cdot\invOp\ul{\calX}  \right ].
}
\end{appendix}
\section*{Acknowledgements}
B. Seibold wishes to acknowledge support by the National Science Foundation through grants DMS--0813648, DMS--1007899 and DMS--1115278. This research was supported in part by the National Science Foundation through major research instrumentation grant number CNS--09--58854.

\nocite{*}
\bibliographystyle{siam}
\bibliography{literature}

\begin{thebibliography}{1}

\bibitem{BraLar00}
{\sc P.~Brantley and E.~W. Larsen}, {\em The simplified {P}$_3$ approximation},
  Nucl. Sci. Eng., 134 (2000), pp.~1--21.

\bibitem{PARCS}
{\sc T.~Downar, D.~Lee, Y.~Xu, and T.~Kozlowski}, {\em {PARCS} v2.6, {U.S. NRC}
  {C}ore {N}eutronics {S}imulator}, School of Nuclear Engineering, {P}urdue
  {U}niversity, draft~ed., 2004.
\newblock https://engineering.purdue.edu/PARCS.

\bibitem{FraKlaLar07}
{\sc M.~Frank, A.~Klar, E.~W. Larsen, and S.~Yasuda}, {\em Time-dependent
  simplified {P$_N$} approximation to the equations of radiative transfer}, J.
  Comput. Phys., 226 (2007), pp.~2289--2305.

\bibitem{Gel60}
{\sc E.~M. Gelbard}, {\em Applications of spherical harmonics method to reactor
  problems}, Tech. Rep. WAPD-BT-20, Bettis Atomic Power Laboratory, 1960.

\bibitem{RohGru96}
{\sc U.~Grundmann and U.~Rohde}, {\em {DYN3D} - {A} 3-dimensional core model
  for steady state and transient analysis in thermal reactors}, in {PHYSOR} 96
  - Breakthrough of Nuclear Energy by Reactor Physics, 1996, pp.~70--79.

\bibitem{LarMorMcG93}
{\sc E.~W. Larsen, J.~Morel, and J.~McGhee}, {\em Asymptotic derivation of the
  simplified {P$_N$} equations}, Proc. ANS Topical Meeting, Mathematical
  Methods and Supercomputing in Nuclear Applications, 1 (1993), pp.~718--730.

\bibitem{Pom93}
{\sc G.~Pomraning}, {\em Asymptotic and variational derivations of the
  simplified {P$_N$} equations}, Ann. Nucl. Energy, 20 (1993), pp.~623--637.

\end{thebibliography}


\begin{thebibliography}{10}

\bibitem{BacBilRibVoi04}
{\sc R.~Backofen, T.~Bilz, A.~Ribalta, and A.~Voigt}, {\em
  {SP$_N$}-approximations of internal radiation in crystal growth of optical
  materials}, Journal of Crystal Growth, 266 (2004), pp.~264--270.

\bibitem{BecGru08}
{\sc C.~Beckert and U.~Grundmann}, {\em Development and verification of a nodal
  approach for solving the multigroup {SP}$_3$ equations}, Ann. Nucl. Energy,
  35 (2008), pp.~75--86.

\bibitem{BraLar00}
{\sc P.~Brantley and E.~W. Larsen}, {\em The simplified {P}$_3$ approximation},
  Nucl. Sci. Eng., 134 (2000), pp.~1--21.

\bibitem{Bru02}
{\sc T.~A. Brunner}, {\em Forms of approximate radiation transport}, Tech. Rep.
  SAND2002-1778, Sandia National Laboratories, 2002.

\bibitem{Cha50}
{\sc S.~Chandrasekhar}, {\em Radiative transfer}, Oxford University Press,
  Dover, 1950.

\bibitem{Chi11}
{\sc G.~Chiba}, {\em Application of the hierarchical domain decomposition
  boundary element method to the simplified {P}$_3$ equation}, Ann. Nucl.
  Energy, 38 (2011), pp.~1033--1038.

\bibitem{Dav57}
{\sc B.~Davison}, {\em Neutron Transport Theory}, Oxford Clarendon Press, 1957.

\bibitem{DenMcC11}
{\sc J.~D. Densmore and R.~G. McClarren}, {\em Moment analysis of angular
  approximation methods for time-dependent radiation transport}, Transport
  Theor. Stat. Phys., 39 (2011), pp.~192--233.

\bibitem{PARCS}
{\sc T.~Downar, D.~Lee, Y.~Xu, and T.~Kozlowski}, {\em {PARCS} v2.6, {U.S. NRC}
  {C}ore {N}eutronics {S}imulator}, School of Nuclear Engineering, {P}urdue
  {U}niversity, draft~ed., 2004.
\newblock https://engineering.purdue.edu/PARCS.

\bibitem{FraKlaLar07}
{\sc M.~Frank, A.~Klar, E.~W. Larsen, and S.~Yasuda}, {\em Time-dependent
  simplified {P$_N$} approximation to the equations of radiative transfer}, J.
  Comput. Phys., 226 (2007), pp.~2289--2305.

\bibitem{FraKlaPin11}
{\sc M.~Frank, A.~Klar, and R.~Pinnau}, {\em Optimal control of glass cooling
  using simplified pn theory}, Transport Theor. Stat. Phys., 39 (2011),
  pp.~282--311.

\bibitem{FraLanSch11}
{\sc M.~Frank, J.~Lang, and M.~Sch\"afer}, {\em Adaptive finite element
  simulation of the time-dependent simplified {P$_N$} equations}, J. Sci.
  Comput., 49 (2011), pp.~332--350.

\bibitem{SeiFra11}
{\sc M.~Frank and B.~Seibold}, {\em Optimal prediction for radiative transfer:
  A new perspective on moment closure}, Kinet. Relat. Models., 4 (2011),
  pp.~717--733.

\bibitem{Gan08}
{\sc B.~D. Ganapol}, {\em Analytical Benchmarks for Nuclear Engineering
  Applications: Case Studies in Neutron Transport Theory}, OECD Nuclear Energy
  Agency, 2008.

\bibitem{Gel60}
{\sc E.~M. Gelbard}, {\em Applications of spherical harmonics method to reactor
  problems}, Tech. Rep. WAPD-BT-20, Bettis Atomic Power Laboratory, 1960.

\bibitem{Gel61}
\leavevmode\vrule height 2pt depth -1.6pt width 23pt, {\em Simplified spherical
  harmonics equations and their use in shielding problems}, Tech. Rep.
  WAPD-T-1182, Bettis Atomic Power Laboratory, 1961.

\bibitem{Gel62}
\leavevmode\vrule height 2pt depth -1.6pt width 23pt, {\em Applications of the
  simplified spherical harmonics equations in spherical geometry}, Tech. Rep.
  WAPD-TM-294, Bettis Atomic Power Laboratory, 1962.

\bibitem{RohGru96}
{\sc U.~Grundmann and U.~Rohde}, {\em {DYN3D} - {A} 3-dimensional core model
  for steady state and transient analysis in thermal reactors}, in {PHYSOR} 96
  - Breakthrough of Nuclear Energy by Reactor Physics, 1996, pp.~70--79.

\bibitem{HabMat75}
{\sc G.~Habetler and B.~Matkowsky}, {\em Uniform asymptotic expansions in
  transport theory with small mean free paths}, J. Math. Phys., 16 (1975),
  pp.~846--854.

\bibitem{HauMcC09}
{\sc C.~Hauck and R.~G. McClarren}, {\em Positive {P$_N$} closures}, Tech. Rep.
  LA-UR 09-03865, Los Alamos Report, 2009.

\bibitem{Jea17}
{\sc J.~H. Jeans}, {\em The equations of radiative transfer of energy}, Monthly
  Notices Royal Astronomical Society, 78 (1917), pp.~28--36.

\bibitem{KloLar06}
{\sc A.~D. Klose and E.~W. Larsen}, {\em Light transport in biological tissue
  based on the simplified spherical harmonics equations}, J. Comput. Phys., 220
  (2006), pp.~441--470.

\bibitem{KloPoe11}
{\sc A.~D. Klose and T.~P\"oschinger}, {\em Excitation-resolved fluorescence
  tomography with simplified spherical harmonics equations}, Phys. Med. Biol.,
  56 (2011), pp.~1443--1469.

\bibitem{Kot07}
{\sc P.~Kotiluoto}, {\em Adaptive tree multigrids and simplified spherical
  harmonics approximation in deterministic neutral and charged particle
  transport}, PhD thesis, University of Helsinki, VTT Technical Research Centre
  of Finland, 2007.

\bibitem{Lar80a}
{\sc E.~W. Larsen}, {\em Diffusion theory as an asymptotic limit of transport
  theory for nearly critical systems with small mean free paths}, Ann. Nucl.
  Energy, 7 (1980), pp.~249--255.

\bibitem{Lar11a}
\leavevmode\vrule height 2pt depth -1.6pt width 23pt, {\em Asymptotic diffusion
  and simplified {P$_N$} approximations for diffusive and deep penetration
  problems. part 1: Theory}, Transport Theor. Stat. Phys., 39 (2011),
  pp.~110--163.

\bibitem{LarKel74}
{\sc E.~W. Larsen and J.~Keller}, {\em Asymptotic solution of neutron transport
  problems for small mean free paths}, J. Math. Phys., 15 (1974), pp.~75--81.

\bibitem{LarMorMcG93}
{\sc E.~W. Larsen, J.~Morel, and J.~McGhee}, {\em Asymptotic derivation of the
  simplified {P$_N$} equations}, Proc. ANS Topical Meeting, Mathematical
  Methods and Supercomputing in Nuclear Applications, 1 (1993), pp.~718--730.

\bibitem{LarMorMcG96}
{\sc E.~W. Larsen, J.~E. Morel, and J.~M. McGhee}, {\em Asymptotic derivation
  of the multigroup {$P_1$} and simplified {$P_N$} equations with anisotropic
  scattering}, Nucl. Sci. Eng., 123 (1996), p.~328.

\bibitem{LarThoKlaSeaGoe02}
{\sc E.~W. Larsen, G.~Th\"ommes, A.~Klar, M.~Sea\"id, and T.~G\"otz}, {\em
  Simplified {P$_N$} approximations to the equations of radiative heat transfer
  and applications}, J. Comput. Phys., 183 (2002), pp.~652--675.

\bibitem{McC11}
{\sc R.~G. McClarren}, {\em Theoretical aspects of the simplified {P$_N$}
  equations}, Transport Theor. Stat. Phys., 39 (2011), pp.~73--109.

\bibitem{MorMcGLar96}
{\sc J.~Morel, J.~McGhee, and E.~W. Larsen}, {\em A three-dimensional
  time-dependent unstructured tetrahedral-mesh {SP$_N$} method}, Nucl. Sci.
  Eng., 123 (1996), pp.~319--327.

\bibitem{Pom93}
{\sc G.~Pomraning}, {\em Asymptotic and variational derivations of the
  simplified {P$_N$} equations}, Ann. Nucl. Energy, 20 (1993), pp.~623--637.

\bibitem{SalKnu00}
{\sc K.~Salari and P.~Knupp}, {\em Code verification by the method of
  manufactured solutions}, Tech. Rep. SAND2000-1444, Sandia National
  Laboratories, 2000.

\bibitem{SchSeaJanKla05}
{\sc E.~Schneider, M.~Sea\"id, J.~Janicka, and A.~Klar}, {\em Validation of
  simplified {P$_N$} models for radiative transfer in combustion systems},
  Commun. Numer. Meth. Engng., 24 (2008), pp.~85--96.

\bibitem{SeaFraKla04}
{\sc M.~Sea\"id, M.~Frank, A.~Klar, R.~Pinnau, and G.~Th\"ommes}, {\em
  Efficient numerical methods for radiation in gas turbines}, J. Comp. Applied
  Math., 170 (2004), pp.~217--239.

\bibitem{SeiboldFrank2009}
{\sc B.~Seibold and M.~Frank}, {\em Optimal prediction for moment models:
  {C}rescendo diffusion and reordered equations}, Contin. Mech. Thermodyn., 21
  (2009), pp.~511--527.

\bibitem{SeiboldFrank2012}
\leavevmode\vrule height 2pt depth -1.6pt width 23pt, {\em \texttt{StaRMAp} --
  {A} {S}taggered grid {R}adiation {M}oment {Ap}proximation solver}, in
  preparation,  (2012).

\bibitem{ThoSta99}
{\sc G.~Thomas and K.~Stamnes}, {\em Radiative Transfer in the Atmosphere and
  Ocean}, Cambridge University Press, Cambridge, 1999.

\bibitem{TomLar96}
{\sc D.~I. Tomasevic and E.~W. Larsen}, {\em The simplified {P}$_2$
  approximation}, Nucl. Sci. Eng., 122 (1996), pp.~309--325.

\bibitem{ZheHan11}
{\sc H.~Zheng and W.~Han}, {\em On simplified spherical harmonics equations for
  the radiative transfer equation}, J. Math. Chem., 49 (2011), pp.~1785--1797.

\end{thebibliography}

\end{document}